\documentclass[MNA]{mna} %

\usepackage{stmaryrd}
\usepackage{xfrac}
\usepackage{bm}
\usepackage{booktabs}

\usepackage{url}
\usepackage{breakurl}

\usepackage{orcidlink}

\newcommand\blfootnote[1]{%
	\begingroup
	\renewcommand\thefootnote{}\footnote{#1}%
	\addtocounter{footnote}{-1}%
	\endgroup
}

\SHORTTITLE{The geometry of color space}

\TITLE{Perceptual spaces and their symmetries: \\ The geometry of color space}

\AUTHORS{%
	Nicolás~Vattuone \orcidlink{0000-0002-3478-1154} $^{1, 2}$ \and 
	Thomas~Wachtler \orcidlink{0000-0003-2015-6590} $^1$
	\and
		Ines~Samengo 
		\orcidlink{0000-0002-5241-3697}
		$^2$}
	
\SHORTAUTHOR{N.~Vattuone, T.~Wachtler and I.~Samengo }


\AMSSUBJ{91E30} 

\SUBMITTED{January 19, 2021} 
\ACCEPTED{June 24, 2021} 

\ARXIVID{2010.00541} 

\VOLUME{1}
\YEAR{2021}
\PAPERNUM{1}
\DOI{10.46298/mna.7108}

\ABSTRACT{Our sensory systems transform external signals into neural activity, from which percepts are produced. We are endowed with an intuitive notion of similarity between percepts, that need not reflect the proximity of the physical properties of the corresponding external stimuli. The quantitative characterization of the geometry of percepts is therefore an endeavour that must be accomplished behaviorally. Here we characterized the geometry of color space using discrimination and matching experiments. We proposed an individually tailored metric defined in terms of the minimal chromatic difference required for each observer to differentiate a stimulus from its surround. Next, we showed that this perceptual metric was particularly adequate to describe two additional experiments, since it revealed the natural  symmetry of perceptual computations. In one of the experiments, observers were required to discriminate two stimuli surrounded by a chromaticity that differed from that of the tested stimuli. In the perceptual coordinates, the change in discrimination thresholds induced by the surround followed a simple law that only depended on the perceptual distance between the surround and each of the two compared stimuli. In the other experiment, subjects were asked to match the color of two stimuli surrounded by two different chromaticities. Again, in the perceptual coordinates the induction effect produced by surrounds followed a simple, symmetric law. We conclude that the individually-tailored notion of perceptual distance reveals the symmetry of the laws governing perceptual computations.}

\begin{document}

\blfootnote{$^1$ Department of 
	Biology II, Ludwig-Maximilians-Universität München and Bernstein Center for 
	Computational 
	Neuroscience, Munich, Germany. Email 
			\texttt{\href{mailto:wachtler@biologie.uni-muenchen.de}{wachtler@biologie.uni-muenchen.de}}}
\blfootnote{$^2$ Department of Medical Physics and Instituto Balseiro, Centro 
Atómico Bariloche, Argentina. \\ Email
\texttt{\href{mailto:nicolas.vattuone@ib.edu.ar}{nicolas.vattuone@ib.edu.ar}}  and
\texttt{\href{mailto:ines.samengo@ib.edu.ar}{ines.samengo@ib.edu.ar}}}



\section{Introduction}
\label{s:introduction}

The neural computations involved in conscious perception, reflections about the world, and action planning, are not performed on external stimuli, but rather on our internal representations of those stimuli. To execute such computations, we are endowed with an intuitive notion of similarity between stimuli. For example, we can typically tell whether two faces are alike or not, whether two tools are exchangeable, or whether two colors are more or less similar. This ability suggests that percepts can be modeled as elements of an abstract space equipped with a notion of distance, so that similar objects be close to each other. If percepts and neural representations  are governed by regularities of the natural world
\cite{Helmholtz1896,Shepard1994,Barlow2001},
the geometry of the perceptual space can be expected to be related with features of the sensory environment and with the specific code with which neurons represent such features.  The first step to characterize such relation is to have a consistent description of perceptual spaces. This is the goal of this paper.

The notion of conceptual spaces has been explored by several studies recently, proposing that, for example, the entorhinal-hippocampal network represents not only spatial information, but more generally, abstract cognitive spaces. A whole variety of cognitive spaces have been investigated, ranging from simple attributes of sensory stimuli \cite{Aronov2017,Radvanski2018}, to highly complex notions, as bird shape \cite{Constantinescu2016} or social hierarchy \cite{Kumaran2016}. The hypothesis is that the items represented in these spaces satisfy geometric constraints such as betweenness and equidistance, so that properties and concepts occupy convex regions \cite{Bellmund2018}. The geometric aspects of physical space are thus attributed to other spaces, and are conjectured to be functionally relevant to guide imagination \cite{Horner2018,Bellmund2016} and decision making \cite{Kaplan2017}.

We access the elements of our conceptual spaces introspectively: We know what color is because we experience color, and the same type of private insight is used to determine notions of similarity. The ontological status of conceptual spaces and their notions of similarity is therefore debatable. Do they have precise properties, and if yes, can we access them objectively? How far can we go? A priori, the existence of an intuitive notion of similarity between a collection of items does not guarantee that the items be describable as points in a space endowed with a topology or a geometry, let alone a Riemannian geometry, in which distances and angles obey exact mathematical relations, and surfaces or volumes can be measured quantitatively. In this paper, we aim at providing an experimental assessment of the existence of a proper geometry. Following \cite{Resnikoff1974}, we work specifically with the space of colors, although the procedure is also valid for other conceptual spaces. 

We first assume that colors form a manifold, that is, a  topological space that can be locally and smoothly mapped to $\mathbb{R}^n$. This assumption is grounded on the observation that any pair of colors can be connected by a trajectory containing items whose physical and perceptual attributes vary continuously. By definition, a differential manifold comes with an Atlas, that is, the set of all the possible coordinate charts over the manifold. Any such system of coordinates may be employed to parametrize the chromaticity of visual stimuli, and the choice does not alter the perceived color. Throughout this paper, the word ``color'' is used for the percept (the private experience), and the word ``chromaticity'', for the physical properties of the electromagnetic spectrum humans are sensitive to. 
For most observers, three coordinates suffice to describe uniform chromatic stimuli. 

Several notions of distance have been proposed in the literature, using criteria based on Weber-Fechner's law \cite{Helmholtz1896,Schrodinger1920,Stiles1946} or on the premise that color space is homogeneous under the group of linear transformations \cite{Resnikoff1974, Provenzi2020, Berthier2019, Berthier2020}. Our emphasis here is twofold: To search for a notion of distance that (1) describes chromatic perceptual effects in the simplest possible manner, and (2) is applicable to multiple perceptual paradigms involving chromatic stimuli.

The search for simplicity is not just for operational convenience. To make an analogy with physics, one and the same physical law can look extremely simple or extremely complicated, depending on the metric we choose for space and time. Space-time itself is intangible, we only have access to events, ultimately perceived as sensory experience. Yet, space and time reveal themselves in the model we construct of the world around us. We observe regularities in the world, and we are able to predict (some of) those regularities by assuming that events take place at a particular place at a particular time, and that they are governed by the laws of physics. Quite remarkably, these laws (be they in their intuitive form, or in their mathematical formulation) become particularly simple when space and time are measured in specific systems: inertial and cartesian. In these systems, classical physics is isotropic and homogeneous, so all equations -- Newton's Mechanics, Coulomb's Law, Maxwell Field Equations, Heat equation, etc. -- only depend on the relative distances between particles, and remain invariant to rigid translations and rotations. Predictability would be seriously challenged if the form of these equations evolved as time went by, or as we moved from one place to the other. Euclidean inertial systems, hence, play a very special role in our mental representation of events. In this paper, we pose the question whether a similar situation can be claimed of specific spaces of phenomenal experience. The geometry of color space is itself also intangible. Yet, a metric of color space may exist, in which perceptual chromatic effects appear to be homogeneous and isotropic. Importantly, in the quest for a privileged metric, several 
perceptual paradigms need to be considered, and this is why the second goal is also required. A metric that provides a simple description of only a single experiment cannot be claimed to characterize color per se; it is more a property of one specific task in which chromatic information intervenes.

Assuming such metric exists, the resulting geometry may or may not be Riemannian, that is, it may or may not result from a metric tensor. If it does, then the set of points that are all at the same (infinitesimal) distance from a given chosen point conform an ellipsoid, from which the metric tensor can be derived. In the vicinity of each point, a special coordinate system known as the ``normal coordinates'' exists, in which the first derivatives of the metric vanish, making the geometry locally flat. In General Relativity, the existence of normal coordinates is the mathematical formulation of the equivalence principle, stating that in a free falling - or inertial - system, spacetime is locally flat. If, additionally, the Riemannian manifold has zero curvature, in the normal coordinates the metric tensor becomes the identity everywhere, so the perceptual distance becomes Euclidean. In this paper, the normal coordinates are called ``perceptual'', since they are derived from perceptual experiments. In these coordinates the symmetries of perception are most naturally revealed, just as Euclidean coordinates of physical space are the ones that most simply reveal the symmetries of Newtonian dynamics. 

To illustrate the meaning of the type of perceptual symmetries we are interested in, we briefly describe the effect of chromatic induction, by which a chromatic context surrounding a stimulus modifies the color of the stimulus \cite{Jameson1964,Ware1982,Wachtler2001}. For example, a green stimulus appears yellowish when surrounded by cyan, and bluish when surrounded by orange. This effect implies that the function that transforms the activities of photoreceptors into a higher-level representation of color depends on the chromaticity of the surround. The perceptual shift is repulsive, since the presence of the surround shifts the perceived stimulus color in color space away from that of the surround \cite{Eichengreen1976, Ware1982, Smith1996, Wachtler2001, Ekroll2004, Hansen2007}. The shift is also non-uniform, since its magnitude, when reported in any of the color coordinates normally used in colorimetry, varies from location to location in color space \cite{Klauke2015}. One can then ask whether a coordinate transformation exists that makes this effect isotropic and homogeneous throughout color space, and thereby, more \emph{symmetric}.

Chromatic induction is not the only perceptual effect revealing the inhomogeneity that color space appears to have in the usually employed coordinate systems. An alternative example is the fact that just-noticeable differences obtained in discrimination tasks vary throughout color space \cite{MacAdam1942, Witzel2013}. Several studies have posed the question whether a coordinate transformation exists that makes the just-noticeable differences uniform \cite{CIE1971, Hunter1987, Fonseca2018}. So far, there is no reason to believe that the coordinates that make discrimination experiments homogeneous and isotropic are the same as those that make induction phenomena homogeneous and isotropic. However, empirically there are similarites between the anisotropies in color discrimination \cite{Witzel2013} and the anisotropies in color induction \cite{Klauke2015}. Therefore, in this paper we ask the question whether 
\emph{all} inhomogeneities can be eliminated with an adequate choice of the coordinate system. If the answer is positive, the metric of color becomes not only a property of a particular experiment, but of color in general. Moreover, it is not only a matter of subjective experience, but also, a latent variable with which all behavioral responses based on chromaticity can be predicted. Since characterizing the inhomogeneities of all perceptual effects is, in practice, an unreachable goal, we here more modestly characterize three perceptual experiments, and agree to scale down the generality of our conclusions accordingly. In the meanwhile,  we may learn something.

The paper is organized as follows. In the Methods section, we describe the behavioral experiments (Sects.~\ref{s:stimuli}-\ref{s:staircase}), and we define the perceptual coordinates in terms of the metric tensor (Sect.~\ref{sec:natural}). Next, the \emph{Results} section starts by describing the correspondence between items in the external world and items in the internal representation (Sect.~\ref{sect:classes}). This step is important, since the mapping need not be one-to-one. In the case of colors, chromatic contexts cause whole collections of external stimuli to be mapped onto single percepts. Once the correspondence is characterized, and the elements of the perceptual space are identified, the geometric structure of percepts is inferred. In order to constrain the search, in Sect.~\ref{sect:esta} we justify from previous experiments the assumption that the space of colors is approximately flat, and we formalize the symmetries that perceptual laws are presumed to adopt when formulated in terms of a sought distance function. The rest of the Results section describes the experiments. To derive the perceptual coordinates in Sect.~\ref{s:exp1} we report the discrimination thresholds along the $S$ and $L - M$ cardinal directions measured in {\sl Experiment I}. A notion of distance is constructed from the obtained thresholds, individually tailored for each observer. The distance is later employed in {\sl Experiments II} and {\sl III} to provide symmetric descriptions of additional perceptual effects. For example, in Sect.~\ref{s:exp2} ({\sl Experiment II}) we show how the thresholds are modified by chromatic surrounds, and confirm the homgeneity and isotropy of the effect when expressed in the perceptual coordinates. Next, in Sect.~\ref{s:exp3} ({\sl Experiment III}) we report the color shifts induced by chromatic surrounds by performing asymmetric matching experiments, in which the colors to be matched were surrounded by different chromaticities. The shifts can be modeled as the consequence of a repulsive field that, in the perceptual coordinates, is isotropic around the surround color. Importantly, in the last two experiments, the perceptual coordinates are the ones obtained from {\sl Experiment I}, with no additional fitting nor manipulation. Therefore, the results of all three experiments become symmetric in the same coordinate system.  We conclude that percepts such as colors, though belonging to the realm of subjective experience, may exhibit elegant mathematical symmetries when described in the adequate coordinates and with the proper geometry.


\section{Methods}

\subsection{Stimuli}
\label{s:stimuli}

Stimuli were displayed on a 21-inch Sony GDM F$520$ CRT screen, controlled by an 8-bit ATI Radeon HD $4200$ graphics card. The spatial resolution was $1280 \times 1024$ pixels and the refresh rate $85$ Hz. The display was calibrated using a PhotoResearch (Chatsworth, CA)  PR-655 spectroradiometer controlled by the IRIS software \cite{Kellner2016}. Photoreceptor excitations $(\bar{S},\bar{M},\bar{L})$ of a given stimulus were obtained by linearly filtering the stimulus spectrum with the \cite{Stockman2000} cone fundamentals. To define the stimuli, a neutral gray was chosen as reference (luminance = $105~\mathrm{cd/m}^2$, CIE[$x,y$]= $[0.328,0.328]$), with coordinates $\left(\bar{S}_g,\bar{M}_g, \bar{L}_g \right) = (1.48,~ 40.9,~ 75.1)$. The cone contrast coordinates of a stimulus were defined as 
\[
\left(S,M,L\right) = \left( \frac{\bar{S}-\bar{S}_g}{\bar{S}_g} ,\frac{\bar{M}-\bar{M}_g}{\bar{M}_g}  ,\frac{\bar{L}-\bar{L}_g}{\bar{L}_g}\right).
\]
These coordinates are invariant under scaling of each of the cone fundamentals. 
Each pixel of the screen was colored with integer RGB coordinates in the range between 0 and 255. To increase chromatic resolution, additional RGB values representing intermediate chromaticities would be required. Therefore, stimulus patches were filled with pixels of randomized integer coordinates whose values differed at most in one unit, thereby creating a finely dithering pattern with a mean chromaticity corresponding to fractional RGB values. Each pixel was colored with one of the four integer triplets $(RGB)^1, (RGB)^2, (RGB)^3$ and $(RGB)^4$ that were closest to the target fractional $RGB$ value. The four options were chosen in appropriate proportions so that the weighted average was equal to the desired fractional $RGB$ triplet. 
As neighboring colors were indistinguishable at the resolution of single pixels, the resulting stimulus patches appeared uniform to subjects.

All measurements were performed along the two cardinal chromatic axes (Fig.~\ref{f1}C):  the $S$ axis, here denoted as $x_1$ and defined by the condition $L = M = 0$, and the $L - M$ axis $(x_2)$, defined by the conditions $S=0$ and $L+M=0$. 

\subsection{Subjects}
\label{s:subjects}

Seven subjects (4 female, 3 male), aged between 22 and 32 participated in the experiments. Subjects gave written consent for participation. Three of the subjects were informed about the purpose of the study and performed measurements along both cardinal color space axes. The remaining four were na\"{i}ve with respect to the study and performed measurements along a single cardinal  color space axis each, either $x_1 = S$ or $x_2 = L - M$. All observers had normal color vision as assessed by the Farnsworth–Munsell 100 Hue test, and had normal or corrected to normal visual acuity.

\subsection{Procedure}
\label{s:procedure}

This section describes the experimental procedure. The experiments were performed in a darkened room. Subjects were seated and viewed the display from a distance of 90 cm. The size of the screen was  $40 \times 30$ cm, subtending a solid angle of $25^{\circ}$x$19^{\circ}$. Subjects were instructed to fix their gaze on a black circle displayed at the center of the screen. Each experiment began with at least 2 min of adaptation to the lighting conditions, during which the subject received instructions and performed test trials that were not included in the analysis. 


\subsubsection{Experiments {\sl I} and {\sl II}: Discrimination}
\label{s:methods:discrimination}
 
Experiments {\sl I} and {\sl II} determined the minimal chromatic difference that stimuli need to bear in order for an observer to identify them as different. The task for the observer was to detect the one out of four stimuli that was chromatically different from the other three.  

A session consisted of $300$ trials, lasting for approximately $10$ minutes. Throughout a session, the chromaticity $\bm{b}$ of the surround remained fixed
and constantly displayed. At the beginning of each trial, a black circle appeared as a fixation point at the center of the screen. After $500$ ms, four $2^{\circ}$ square patches were displayed for $150$ ms at a center-to-center distance of 2° from the fixation point along the cardinal directions.
Three of the patches were colored with the test chromaticity $\bm{x}$. The fourth patch was the target patch and had a slightly different chromaticity $\tilde{\bm{x}}$. The location of the target patch was varied randomly from trial to trial among the four alternatives. The observer was required to report its position using arrow keys on a keyboard. Subjects had unlimited time to respond. They were allowed to freely set the pace of the experiment by triggering each trial with a key on the keyboard. In each session, the tested chromaticity $\bm{x}$ remained fixed, and the altered chromaticity $\tilde{\bm{x}}$ was chosen randomly among 15 alternatives around $\bm{x}$, each sampled $20$ times. 

\begin{figure}[ht]
\centering
\includegraphics[scale=0.5]{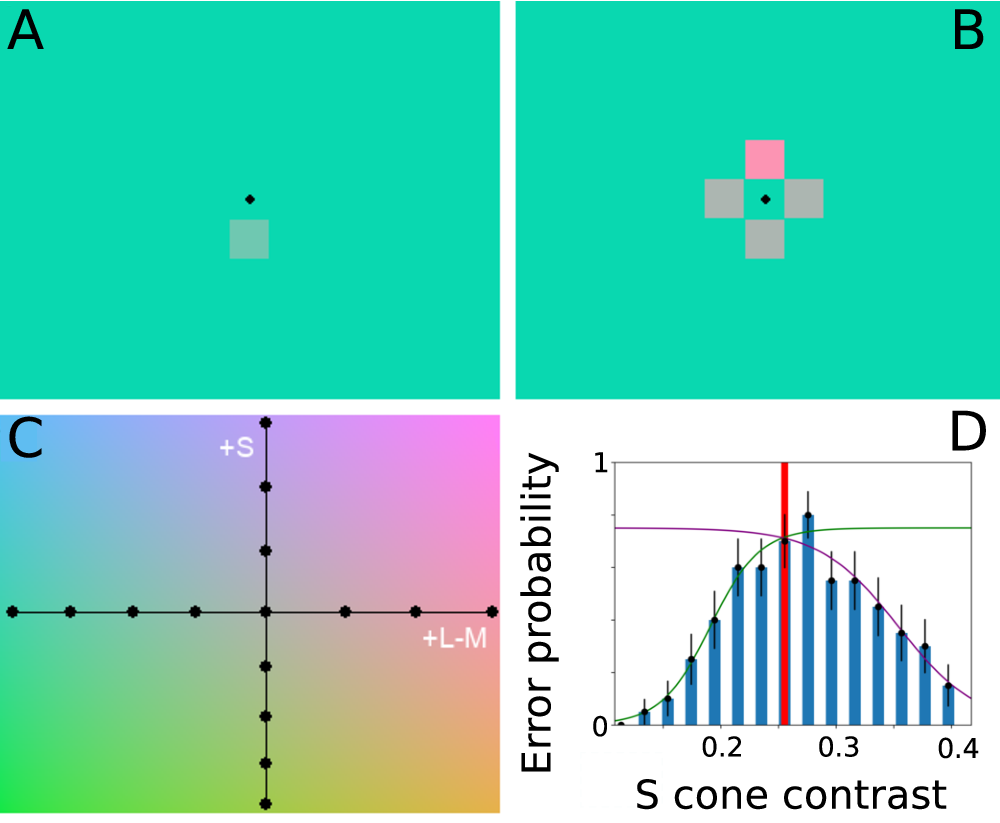}
\caption{{\bf Experimental paradigms of the discrimination experiments}. ({\sl A, B}): Stimulus displays for discrimination experiments, performed with surround chromaticity $\bm{b}$ equal to ({\sl A}) or different from ({\sl B}) the tested chromaticity $\bm{x}$. {\sl C}: Thresholds were measured for eight tested chromaticities on each axis (black circles). On the axis $S$ the cone-contrast values used for Experiment I were$\lbrace -0.58, -0.46, -0.33, -0.18, 0, 0.16,  0.35,  0.54 \rbrace$, and for axis $L-M$ the values were:$\lbrace -0.17, -0.13 , -0.09, -0.05, -0.01 ,  0.04, 0.08, 0.12 \rbrace$.The intersection of the axes corresponds to the reference gray $(0,0)$. {\sl D}: Error probability (black bars) reported by subject S2 in a session of $N = 20$ trials per target stimulus, as a function of the $S$ cone contrast $\tilde{x}_1$ of the altered stimulus, for a fixed tested stimulus $x_1$ (red bar). Random responses are expected to produce 75\% of incorrect identifications. As the difference $|\tilde{x}_1 - x_1|$ between the dissimilar patch and the other three patches increases, the error probability drops. Error bars denote the standard errors for corresponding binomial distributions. The fitted parameters of Eq.~\ref{eq:asimetria} are $a_\ell = 0.043 \pm 0.004, b_\ell = 0.191 \pm 0.003, a_r = 0.07 \pm 0.01, b_r = 0.354 \pm 0.006$.} \label{f1}
\end{figure}

In {\sl Experiment I}, the chromaticity of the surround coincided with that of the three test patches (Fig.~ \ref{f1}A), so the
observer had to detect the location of the target patch in a uniform surround. In {\sl Experiment II}, the surround $\bm{b}$ had a different chromaticity, so the observer had to compare the four patches, and detect the target patch (Fig.~\ref{f1}B). In both experiments, the chromaticity of the surround was varied systematically along the $b_1 = S$ and the $b_2 = L - M$ dimensions, while the luminance $L + M$ was maintained constant (Sect.~\ref{s:stimuli}). In {\sl Experiment I}, each time the surround $\bm{b}$ was modified, the tested chromaticity $\bm{x}$ was changed accordingly. Eight different chromaticities were tested along each axis (values in Fig.~\ref{f3}). In {\sl Experiment II}, colors of stimuli and surrounds were varied independently. Three surround chromaticities were employed on each cardinal axis, with cone contrast coordinates $S = -0.24, 0 , 0.16$, and $L - M = -0.03, 0, 0.03$. Eight different chromaticities were used for the test stimuli on each axis (Fig.~\ref{f1} C). 

As observers selected one among four options, the chance error rate was $75$\%. This percentage diminished with increasing discriminability. Figure ~\ref{f1}{\sl D} displays the error probability for subject S2 in a given session for different altered chromaticities $\tilde{\bm{x}}$  around the tested chromaticity $\bm{x}$. We defined the discrimination threshold $\varepsilon$ as the value of $\tilde{x}$ for which the error probability was equal to the midpoint between pure chance and perfect performance, i.e. when the error probability was $37,5$\%. Thresholds may be different for increasing and decreasing cone activation \cite{Chichilnisky1996}, implying that the bar plot of Fig.~\ref{f1}{\sl D} need not be symmetric around the maximum. In order to take asymmetries into account, left-side ($\ell$) and right-side ($r$) thresholds were estimated by separately fitting sigmoid functions to the data for each side of the tested chromaticity. The fitted functions were
\begin{equation} \label{eq:asimetria}
P_{\ell,r}(\tilde{x}) = 0.375\left[ 1  \pm  \tanh \left(  a_{\ell, r} (\tilde{x} -b_{\ell, r} ) \right) \right],
\end{equation}
with fitted parameters $a_{\ell}$ and $b_{\ell}$ or $a_r$ and $b_r$ for the left and right side, respectively. The left (decreasing cone contrast) and right (increasing cone contrast) thresholds of a the reference chromaticity $\bm{x}$ were defined as $\Delta_{\ell, r} = \vert  b_{\ell, r} - x\vert$, and the mean threshold, as $\varepsilon = (\Delta_\ell + \Delta_r)/2$. 


\subsubsection{Experiment 3: Asymmetric matching}
\label{sect:asymmetric}

\begin{figure}[ht!]
\centering
\includegraphics[scale=0.5]{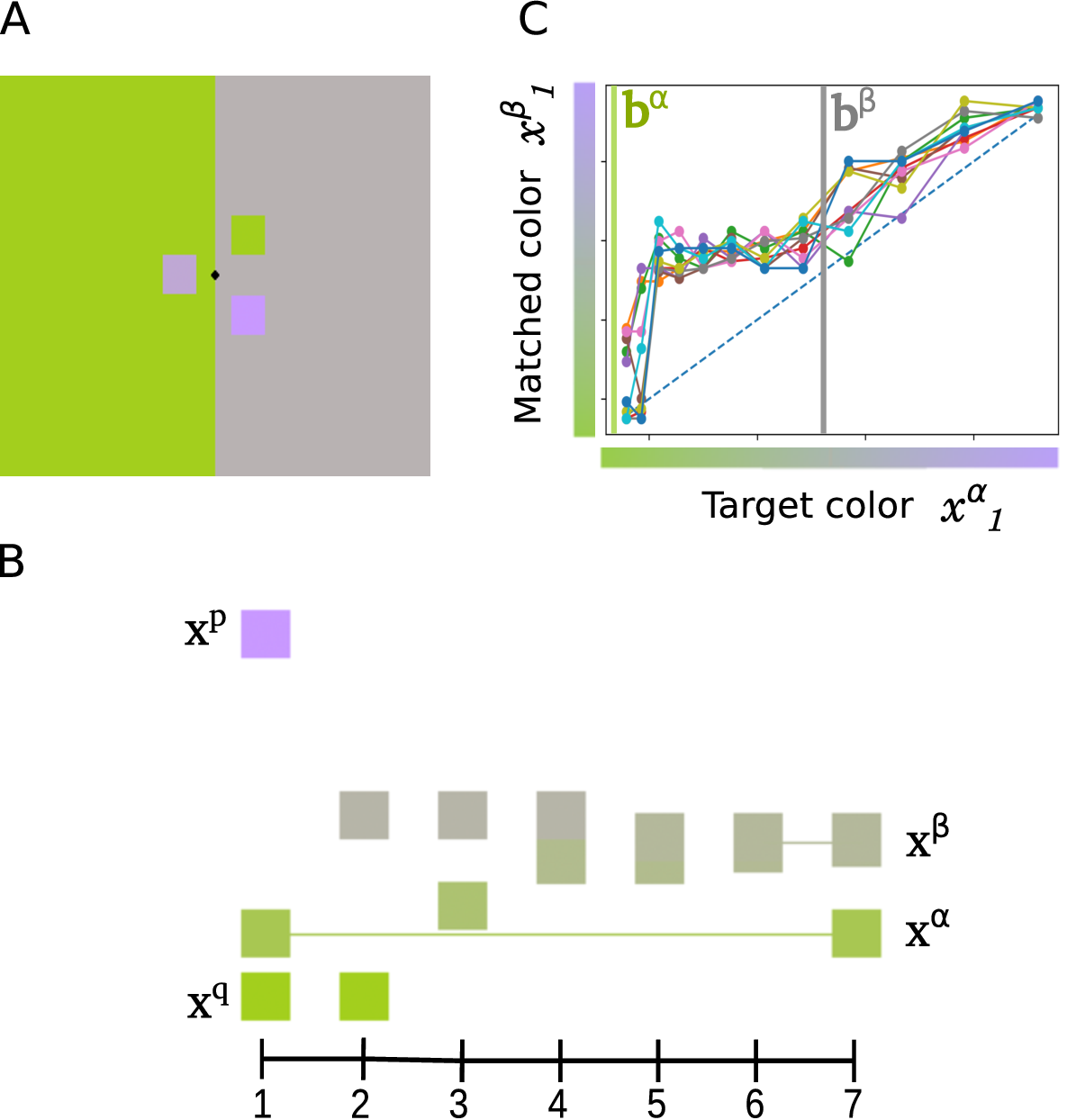}
\caption{{\bf Experimental paradigm of the matching experiments.} {\sl A}: Two patches $\bm{x}^p$ and $\bm{x}^q$ were presented on the right, surrounded by $\bm{b}^\beta$. The observer had to report which of the patches appeared most similar to the target stimulus $\bm{x}^\alpha$ on the left, which was surrounded by $\bm{b}^\alpha$. {\sl B}: Sequence of chromaticities $\bm{x}^p$ and $\bm{x}^q$ appearing in response to the choices of the subject. Horizontal axis: trial sequence. Vertical axis: coordinate $S = x_1$ of each patch. Horizontal line: target color $\bm{x}^\alpha$. The staircase sequence contained $6$ trials, after which the final matched stimulus $\bm{x}^\beta$ was calculated as the average of $\bm{x}^p_{\bm{6}}$ and $\bm{x}^q_{\bm{6}}$. {\sl C}:  Chromaticity $x^{\beta}$  presented on surround $\bm{x}^\beta$ (gray line) that matched the target $\bm{x}^\alpha$ presented on surround $\bm{b}^{\alpha}$ (green line). Different lines represent the converged chromaticity obtained in each of the 10 sequences responded by observer S2.  
\label{f2} }
\end{figure}

Colored surrounds alter the color of a test stimulus \cite{Klauke2015}. To test these influences along the cardinal color axes, we performed {\sl Experiment III}, an asymmetric color matching task. In these experiments, test and match stimuli were displayed in surrounds of different chromaticities. In classical color matching experiments \cite{CIE1932, Guild1932, Stiles1959, Wyszecki2000}, subjects 
performed the match to the test stimulus by adjusting the match stimulus without constraints on fixation or presentation time. To control for these factors, and to work in conditions that were similar to those of the discrimination experiments, we used a forced-choice paradigm. In each trial, the observer was presented two candidate patches on one half of the screen surrounded by chromaticity $\bm{b}^{\beta}$, and was instructed to select among them the one perceived as most similar to the target patch displayed on the other half of the screen, surrounded by chromaticity $\bm{b}^{\alpha}$ (Fig.~\ref{f2}A). The side of the screen occupied by the target patch was randomized in each trial. 

For each combination of test stimulus $\bm{x}^{\alpha}$ in surround $\bm{b}^{\alpha}$, here denoted as $\bm{x}^{\alpha} \sslash \bm{b}^{\alpha}$, the aim was to determine the match $\bm{x}^\beta$ on the surround $\bm{b}^{\beta}$. In other words, we searched for the chromaticity $\bm{x}^\beta$ that fulfilled the perceptual equality $\bm{x}^\beta \sslash \bm{b}^\beta \sim \bm{x}^\alpha \sslash \bm{b}^\alpha$. Here, the symbol ``$\sim$'' means that  stimulus $\bm{x}^\alpha$ surrounded by $\bm{b}^\alpha$ appears to have the same color as stimulus $\bm{x}^\beta$ surrounded by $\bm{b}^\beta$.
The search for $\bm{x}^\beta$ was performed as a staircase procedure (Sect.~\ref{s:staircase}). 

 Three pairs of surrounds were used for each axis. Two of the pairs combined the neutral reference gray corresponding to the origin of color space with the maximally and minimally attainable coordinates on the axis, respectively. The cone contrasts of these surrounds with respect to the neutral gray were $S_{\mathrm{min}}= -0.35$, $S_{\mathrm{max}}=0.25$ for axis S, and  ${L-M}_{\mathrm{min}}=-0.20$, ${L-M}_{\mathrm{max}}=0.15$ for axis L-M. The third pair did not include gray, and contained the two other surrounds of {\sl Experiment II} that were unsaturated colors in cardinal directions. Their cone contrasts with respect to the neutral gray were $S= -0.24$ and $S=0.16$ for axis $S$,  $L-M =-0.03$ and $L-M=0.03$ for axis $L-M$. The first two pairs were useful to assess the shifts produced by fairly saturated colors, and to measure the structure of the induction when the distance between the colored and neutral surround was large. The third pair was selected so as to connect the results of {\sl Experiment III} with those of {\sl Experiment II}, and to assess the behavior of the shift for desaturated surrounds.

Subjects initiated each trial by pressing a key on the keyboard. At the beginning of each presentation both surrounds were shown for $200$ ms, together with a black circle as fixation point. Then, a patch of chromaticity $\bm{x}^\alpha$ was presented on $\bm{b}^\alpha$ and two patches with chromaticities $\bm{x}^p$ and $\bm{x}^q$ appeared against the surround $\bm{b}^\beta$, one above the other (top and bottom locations randomized) for $500$ ms. All patches subtended a visual angle of  $2^\circ$. After the stimulus presentation, a masking stimulus was displayed for $500$ ms, consisting of randomly sized and located square patches with a balanced distribution of colors along the corresponding axis, to reduce afterimages \cite{Wachtler2001}. Then, the uniform neutral gray background was displayed, and the subject was required to respond whether the top or the bottom patch ($\bm{x}^p$ or $\bm{x}^q$) was most similar to $\bm{x}^\alpha$ by pressing the corresponding arrow key on the keyboard.

\subsubsection{Staircase procedure}
\label{s:staircase}

{\sl Experiment III} was structured in sequences, one sequence defined as $6$ consecutive trials. In each trial, two patches with chromaticities $\bm{x}^p$ and $\bm{x}^q$ appeared surrounded by $\bm{b}^\beta$.  The subject's task was to select the patch that appeared to be most similar to the target $\bm{x}^\alpha$, surrounded by $\bm{b}^\alpha$. The two options $\bm{x}^p$ and $\bm{x}^q$ were meant to be an upper and a lower bound for the matched chromaticity $\bm{x}^\beta$, and were updated progressively throughout the trials of the sequence. In the first trial of the sequence, $\bm{x}^p_{\bm{1}}$ and $\bm{x}^q_{\bm{1}}$ took the maximal and minimal values allowed by the display for the corresponding axis. For instance, along the $x_1 = S$ axis, initially $\bm{x}^p_{\bm{1}}$ was a maximally saturated purple and $\bm{x}^q_{\bm{1}}$, a maximally saturated yellow-green. At trial $i$, the subject decided whether $\bm{x}^p_{\bm{i}} \sslash \bm{b}^\beta$ or $\bm{x}^q_{\bm{i}} \sslash \bm{b}^\beta$ was perceived as more similar to $\bm{x}^\alpha \sslash \bm{b}^\alpha$. For trial $i + 1$, the non-selected chromaticity at step $i$ was updated by the midpoint between the two previous options, that is, 
\begin{align*}
\bm{x}^p_{\bm{i + 1}} &= \bm{x}^p_{\bm{i}} - (1- z_i) \ \frac{\bm{x}^p_{\bm{i}} - \bm{x}^q_{\bm{i}}}{2} \\
\bm{x}^q_{\bm{i + 1}} &= \bm{x}^q_{\bm{i}} +  z_i \ \frac{\bm{x}^p_{\bm{i}}-\bm{x}^q_{\bm{i}}}{2}
\end{align*}
where $z_i = 0$ if the subject chose $\bm{x}^p_{\bm{i}}$, and $z_i = 1$, otherwise. Both progressions of chromaticities $\bm{x}^p_{\bm{i}}$ and $\bm{x}^q_{\bm{i}}$ were bounded and monotonic, and their distance decreased exponentially, so they both converged to the same value $\bm{x}^\beta$. We estimated this value as $(\bm{x}^p_{\bm{6}} + \bm{x}^q_{\bm{6}}) / 2$, and interpreted as the color for which $\bm{x}^\beta \sslash \bm{b}^\beta$ matched $\bm{x}^\alpha \sslash \bm{x}^\alpha$. We verified that after $6$ steps, the two bounds were indistinguishable.  

In principle, the obtained $\bm{x}^\beta$ is not guaranteed to be an exact match. Still, the choices of the subject that lead to $\bm{x}^\beta$ are those that minimize the perceptual distance, so $\bm{x}^\beta$ is the stimulus that makes $\bm{x}^\beta \sslash \bm{b}^\beta$ as similar as possible to $\bm{x}^\alpha \sslash \bm{b}^\alpha$, among the available options. This argument is equivalent to the projection notion employed by \cite{Song2019}. In our experiments, exact matches are only possible if the sequence of presented colors actually approaches the target $\bm{x}^\beta$. If $\bm{x}^\beta$ indeed lies along the explored axis, then perfect matches become possible, except perhaps for small discrepancies produced by the different position on the retina excited by the three compared stimuli. Instead, if the target lies outside the explored axis, perfect matches are downright impossible. We verified that for each axis, the induction along the direction that is orthogonal to the tested axis was indistinguishable from noise (data not shown). Therefore, in our paradigm, all the options $\bm{x}^p$ and $\bm{x}^q$ available to the subject belonged to the same cardinal axis connecting $\bm{x}^\alpha, \bm{b}^\alpha$ and $\bm{b}^\beta$.


\subsection{The perceptual coordinates}
\label{sec:natural}

Discrimination thresholds can be understood as the granularity with which the space of colors is perceived. The underlying assumption is that the neural activities involved in representing two colors separated by less than the threshold are not reliably different. The size of thresholds, and their variation throughout color space, depend on the coordinate system. In this paper, we report the experimental results in the cone contrast coordinates $x_1 = S$ and $x_2 = L - M$ \cite{Derrington1984}, maintaining the total luminance $x_3 = L + M$ fixed, as done in previous studies \cite{Klauke2015}. Each color is represented as a column vector $\bm{x}$ with components $x_1$ and $x_2$. Sect.~\ref{sect:classes} explains the connection between the perceived color and the chromaticities of both stimulus and surround. In order to reveal the symmetries of color space, we use the measured thresholds to define a metric tensor $J$, and the perceptual coordinates $(x_1', x_2')$ of each observer. In this section, we show how to  transform from the cone contrasts to the perceptual coordinates. 

The metric tensor $J(\bm{x})$ of the space of colors must be symmetric and non-negative, and it allows us to calculate scalar products $(\bm{v}^\alpha)^t J(\bm{x})\bm{v}^\beta$ between vectors $\bm{v}^\alpha$ and $\bm{v}^\beta$ of the tangent space at $\bm{x}$. In a neighbourhood around $\bm{x}$ and for a certain coordinate chart $\lbrace x_i \rbrace$, vectors can be mapped to small displacements in the space through the flow  of the coordinate vector fields $\lbrace \hat{e}_i  = \frac{ \partial }{ \partial x^i} \rbrace$. In our experiments, by expressing the chromaticity of stimuli in a specific coordinate system, we select the chart we use in the space of colors. For a stimulus that is close to $\bm{x}$, we use the notation $\bm{x} + {\rm d}{\bm x}$. For sufficiently small displacements, the mapping between the chromaticities around $\bm{x}$ and the tangent space at $\bm{x}$ allows us to interpret ${\rm d}{\bm x}$ both as the change in chromaticity and the tangent vector which generates the infinitesimal displacement from $\bm{x}$ to $\bm{x} + {\rm d}{\bm x}$.

%
%
%
The  line element ${\rm d}\ell$ measuring the distance between a given color $\bm{x}$ and the infinitesimally displaced color $\bm{x} + {\rm d}{\bm x}$ is 
\begin{eqnarray} 
{\rm d}\ell &=& {\rm d}(\bm{x}, \bm{x} + {\rm d}\bm{x}) \nonumber \\
&=& \sqrt{{\rm d}\bm{x}^t \ J(\bm{x}) \ {\rm d}\bm{x}} \label{e:1} \\
&=& \sqrt{J(\bm{x})_{11} \ ({\rm d}x_1)^2 + 2 \ J(\bm{x})_{12} \ {\rm d}x_1 \ {\rm d}x_{2}  + J(\bm{x})_{22} \ ({\rm d}x_2)^2}, \nonumber
\label{e:infdistan} 
\end{eqnarray}
where the superscript $t$ represents vector transposition. Our aim is to find the tensor $J(\bm{x})$ that  represents perceptual differences, that is, the one for which the distance ${\rm d}\ell$ of Eq.~\ref{e:infdistan} between two neighboring colors $\bm{x}$ and $\bm{x} + {\rm d}{\bm x}$ captures their behavioral discriminability. If an observer is capable of particularly accurate discrimination between $\bm{x}$ and a slightly displaced color along a direction $\hat{\bm{e}}$, the discrimination threshold must be particularly small in this direction. The smaller the threshold, the more sensitive the observer. 

To construct $J(\bm{x})$, the discrimination threshold between color $\bm{x}$ and a displaced color along the direction $\hat{\bm{e}}$ needs to be measured for every possible direction $\hat{\bm{e}}$. Operationally, this means to move progressively away from $\bm{x}$, in small steps that add up to $\varepsilon$, along the direction $\hat{\bm{e}}$, and to test whether the reached color $\bm{x} + \varepsilon \ \hat{\bm{e}}$ can be discriminated from $\bm{x}$ with a pre-set accuracy. If this is the case, then $\bm{x}$ and $\bm{x} + \varepsilon \hat{\bm{e}}$ are defined to be at a fixed distance from each other. In this paper, we define the units of length by setting this distance as equal to $1$: A length of one unit in color space  yields a threshold error rate of 37.5\% in {\sl Experiment I} (Sect.~\ref{s:methods:discrimination}). If the reached color is discriminated from $\bm{x}$ with a larger error rate, the size of $\varepsilon$ is increased, and the procedure is iterated until the first color below the threshold is reached. 

If thresholds are assumed to vary continuously with the direction $\hat{\bm{e}}$, the lowest-order analytical expression that captures their directional modulation is given by the equation of an ellipse, obtained by setting the distance $d\ell$ of Eq.~\ref{e:infdistan} equal to $1$ and squaring the resulting equality. The vector $(\varepsilon \hat{\bm{e}})^t = (\varepsilon_1, \varepsilon_1)$ is therefore a solution of 
\begin{equation}
(\varepsilon \hat{\bm{e}})^t \ J(\bm{x}) \ \varepsilon \hat{\bm{e}} = 
\begin{pmatrix}  \varepsilon_1 & \varepsilon_2  \end{pmatrix}    \left(\begin{array}{cc}J_{11}(\bm{x}) & J_{12}(\bm{x}) \\ J_{21}(\bm{x}) & J_{22}(\bm{x}) \end{array} \right) \left(\begin{array}{c} \varepsilon_1 \\ \varepsilon_2 \end{array}\right) = 1,
\label{e:jinve}
\end{equation}
which defines an ellipse because of the positive definiteness of $J$. The eigenvectors of $J(\bm{x})$ are aligned with the principal axes of the ellipse, and the eigenvalues are the inverse square of their lengths. An ellipse centered at point $\bm{x}$ is determined by three non-colinear points, or equivalently by the length of its semiaxes and its orientation. Therefore, by measuring the discrimination thresholds along three directions, and using Eq.~\ref{e:jinve}, a system of three equations and three unkowns is obtained, the solution of which are the components of the symmetric tensor $J$.

The length of a path connecting two remote colors is obtained by integrating local increments $\mathrm{d}\ell$ along the trajectory, so the total length is the number of thresholds that need to be crossed to travel from one color to the other. Of course, the metric tensor may vary along the path, and different paths connecting the same pair of points may have different lengths. The distance is then defined as the length of the shortest path. For practical reasons, $J(\bm{x})$ cannot be estimated for the infinite collection of points $\bm{x}$ composing the trajectory. In order to calculate the path integral, hence, $J(\bm{x})$ must be estimated for a subset of colors $\bm{x}$ that sample the curve under study with sufficient resolution. The intermediate tensors are interpolated under the assumption that the discrimination ability varies continuously between samples.

Under adaptation to the surround, the results of \cite{Krauskopf1992} indicated that, in the cone contrast coordinates, the off-diagonal terms of $J(\bm{x})$ vanish. In this case, thresholds only need to be measured along the cardinal axes $\bm{e}^1$ and ${\bm e}^2$. Infinitesimal distances along the axes then read
\begin{eqnarray}
{\rm d}\ell_i &=& {\rm d}\left(\bm{x}, \bm{x} + {\rm d}x_i \ \bm{e}^i\right) \nonumber \\
&=& \sqrt{J_{ii}(\bm{x}) \ ({\rm d}x_i)^2} \nonumber \\
&=& \frac{|{\rm d}x_i|}{\varepsilon(\bm{e}^i)}, \label{e:jepsilon}
\end{eqnarray}
where the subscript $i$ indicates either the $S$ $(i = 1)$ or the $L - M$ $(i = 2)$ coordinate. The distance between two colors $\bm{x}^a$ and $\bm{x}^b = \bm{x}^a + \Delta \bm{e}^i$ that differ by a vector aligned with the cardinal axes $i$ is found by integration
\begin{eqnarray}
{\rm d}\left(\bm{x}^a, \bm{x}^b \right)
&=& \int_{\bm{x}^a}^{\bm{x}^b} {\rm d}\ell \nonumber \\
&=& \int_{\bm{x}^a}^{\bm{x}^b} \sqrt{J_{ii}(\bm{x})} \ |{\rm d}x_i| \label{e:distancia}
\end{eqnarray}

If $J(\bm{x})$ is diagonal, and in addition, the term $J_{ii}(\bm{x})$ only depends on the component $x_i$ (as verified by \cite{Krauskopf1992}), the space of colors has zero curvature. In this case, a coordinate transformation $\bm{x} \to \bm{x}'$ exists, such that the transformed metric is Euclidean. In Euclidean spaces, all geodesics are straight lines, which greatly simplifies the perceptual shift produced by surrounds, as explained below. In the new coordinates, the discrimination ability of the observer is isotropic and homogeneous, that is, all discrimination ellipses become circles, and all circles have the same size. These are the coordinates that most naturally reveal the perceptual abilities of the subject, and are therefore here called the {\em perceptual} coordinates of the observer. It is easy to prove that the function instantiating the transformation to the perceptual coordinates is
\begin{eqnarray}
x'_1(\bm{x}) &=&  d\left((x^0_1, x^0_2)^t, (x_1, x^0_2)^t\right) \nonumber \\
&=& \int_{x_1^0}^{x_1} \sqrt{J(y_1, x_2^0)} \ {\rm d}y_1, \label{e:x1prima} \\
x'_2(\bm{x}) &=&  d\left((x^0_1, x^0_2)^t, (x_1^0, x_2)^t \right) \nonumber \\
&=& \int_{x_2^0}^{x_2} \sqrt{J(x_1^0, y_2)} \ {\rm d}y_2, \label{e:x2prima} 
\end{eqnarray}
where $d(\bm{x}^p, \bm{x}^q)$ is the distance between colors $\bm{x}^p$ and $\bm{x}^q$, and $\bm{x}^0$ is the origin of the new system of coordinates ($\bm{x}'(\bm{x}^0) = \bm{0}$) and may be chosen arbitrarily.


\section{Results}

\subsection{Classes of equivalence in the space of stimuli $\times$ surrounds}

\label{sect:classes}

In this section, we describe the mapping between external stimuli and percepts, with special emphasis on the role of context. For the sake of simplicity, the only aspect of context that matters is chromaticity, all other aspects (as spatial or temporal structure) are kept uniform.  The color with which a stimulus is perceived depends on the spectral properties of both the stimulus and the surround. Mathematically, this means that
\begin{equation} \label{e:perceived}
\text{Perceived \ color} = \mathrm{Function}[\bm{x}, \bm{b}],
\end{equation}
where $\bm{x}$ and $\bm{b}$ represent the stimulus and the surround, respectively. In trichromats, three numbers suffice to characterize the perceivable properties of the 
light spectrum, giving rise to the well-known $3$-dimensional color spaces, such as $LMS$, $RGB$, $XYZ$, or others. Equation~\ref{e:perceived} suggests that, in a center-surround situation with uniform center and uniform surround, $6$ coordinates are required to specify a percept, $3$ for the color of the stimulus and 3 for the surround.

\begin{figure}[ht]
\centering
\includegraphics[scale = 0.7]{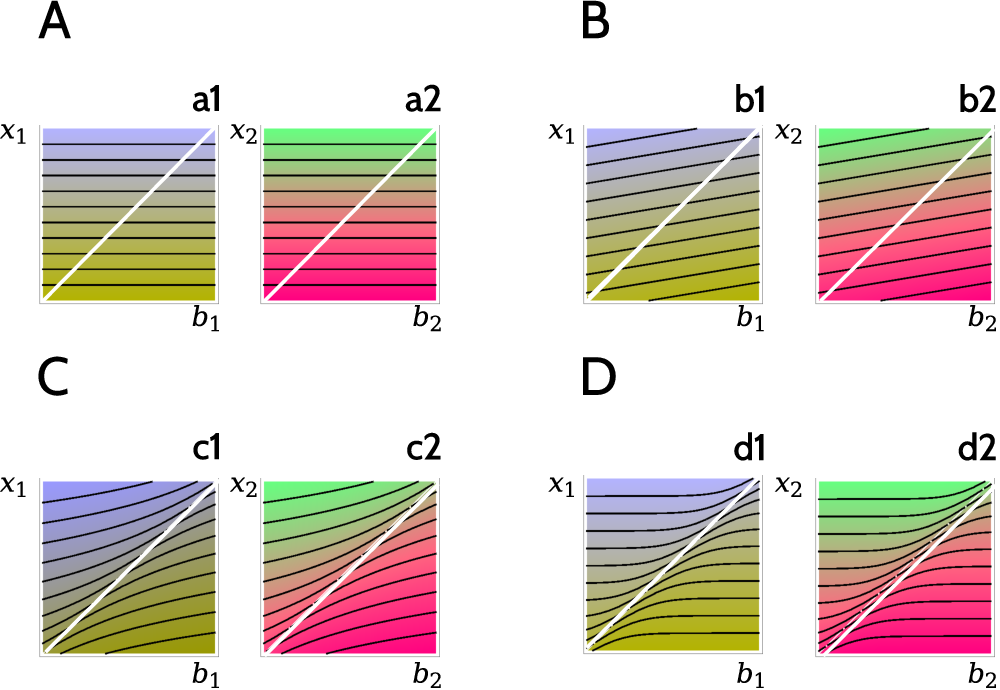}
\caption{{\bf Classes of equivalence}. Four different examples of the structure of the partition induced by classes of equivalence. Black lines represent classes of equivalence, and are obtained by plotting $\bm{\Phi}_{\bm{b}}(\bm{x})$ for fixed $\bm{x}$ (one value per line) and varying $\bm{b}$. The diagonal white line contains the uniform representatives. {\sl A}: The surround does not alter the color of the stimulus, so the classes of equivalence are planar (straight lines). {\sl B}: The surround induces a linear classes of equivalence, as suggested by \cite{Resnikoff1974} and \cite{Provenzi2020}. {\sl C} and {\sl D}: Two other possible partitions of color space, with more complex classes of equivalence.
\label{f0001} } 
\end{figure}

Quite remarkably, although chromatic surrounds modify the way stimuli are perceived, the percept they induce is still a color, since observers engage themselves naturally in asymmetric matching experiments, where they match pairs of stimuli surrounded by different chromaticities. This means that for each stimulus $\bm{x}^\alpha$ presented against surround $\bm{b}^\alpha$, and for each new surround $\bm{b}^\beta$, a new stimulus $\bm{x}^\beta$ can be defined by a function
\begin{equation} \label{e:fi}
    \bm{x}^\beta = \bm{\Phi}_{\bm{b}^\alpha \to \bm{b}^\beta}(\bm{x}^\alpha),
\end{equation}
such that 
\[
    \bm{x}^\alpha \sslash \bm{b}^\alpha \sim \bm{x}^\beta \sslash \bm{b}^\beta.
\]
In asymmetric matching experiments, observers compute the function $\bm{\Phi}_{\bm{b}^\alpha \to \bm{b}^\beta}$. As first noted by \cite{Resnikoff1974}, the matching operation ``$\sim$'' defines an equivalence relation, that is, a relation between pairs of ``$\mathrm{stimulus} \ \sslash \  \mathrm{surround}$'' that is reflexive $(\bm{x} \sslash \bm{b} \sim \bm{x} \sslash \bm{b})$, symmetric (if $\bm{x}^\alpha\sslash \bm{b}^\alpha \sim \bm{x}^\beta \sslash \bm{b}^\beta$ then $\bm{x}^\beta\sslash \bm{b}^\beta \sim \bm{x}^\alpha \sslash \bm{b}^\alpha$), and transitive (if $\bm{x}^\alpha\sslash \bm{b}^\alpha \sim \bm{x}^\beta \sslash \bm{b}^\beta$ and also $\bm{x}^\beta\sslash \bm{b}^\beta \sim \bm{x}^\gamma \sslash \bm{b}^\gamma$, then $\bm{x}^\gamma\sslash \bm{b}^\gamma \sim \bm{x}^\alpha \sslash \bm{b}^\alpha$). All equivalence relations induce a partition in the set they operate upon. In other words, the set of pairs $\bm{x} \sslash \bm{b}$ can be segmented into disjoint subsets, or {\sl classes of equivalence}. All pairs belonging to the same class are pairwise connected with the relation $\sim$, and also, pairs belonging to different classes are not connected with $\sim$. In line with Resnikoff, here we assume that a given {\sl color} is the percept shared by all the pairs that belong to the same class. A color is therefore not a property of a specific stimulus $\bm{x}$, nor even of a specific pair $\bm{x}\sslash \bm{b}$. It is a property of a whole class of pairs. In mathematical terms, color is the quotient space of the original space of pairs and the equivalence relation ``$\sim$''. Therefore, the $6$ coordinates mentioned above constitute a redundant representation of color. Classes of equivalence are $3$-dimensional submanifolds embedded in the $6$-dimensional space defined by stimuli and surrounds. If selecting a color is equivalent to selecting a class, $3$ coordinates suffice. In Fig.~\ref{f0001}, the classes of equivalence are illustrated for four different choices for the function defining the displacements induced by surrounds. Since it is not possible to depict $3$-dimensional submanifolds embedded inside a $6$-dimensional space, the figure shows slices containing the axes $(x_1, b_1)$ and $(x_2, b_2)$, respectively. In these slices, each class appears as a curve.
In Fig.~\ref{f0001}{\sl A}, the surround does not alter the color of the stimulus, and therefore, the classes of equivalence are planar: Irrespective of the surround, $\bm{x} \sslash \bm{b}$ is always perceived the same. In Fig.~\ref{f0001}{\sl B}, classes of equivalence are linear. The surround produces a repulsive effect, which becomes larger as the distance between the surround and the stimulus increases. In panels {\sl C} and {\sl D}, the effect of the surround is more complex.

We now assume that, at least for the unsaturated colors explored in this paper, all equivalence classes contain a unique {\sl uniform representative}, that is, a pair of the form $\bm{x} \sslash \bm{x}$, in which the stimulus coincides with its surround. In Fig.~\ref{f0001}, uniform representatives lie along the white diagonal line, so the assumption means that all classes intersect the diagonal. The hypothesis is supported by the empirical observation that subjects find feasible the task of matching a uniform stimulus $\bm{x} \sslash \bm{x}$ of controlled chromaticity with a target stimulus $\bm{x}'$ presented against a surround of different chromaticity $\bm{b}'$.  In our lab, this feasibility has been verified for the set of target stimuli that can be produced by our computer screen. Although this set does not include maximally saturated colors, it is broad enough to encompass a rich collection of chromaticities. The uniform representative of each class must be unique, since all the members of a class are perceptually indistinguishable, and two uniform representatives of different chromaticity are (by definition of ``different'' ) distinguishable. We define the function $\bm{\Phi}_{\bm{b}}(\bm{x})$ as the one that maps each member $\bm{x} \sslash \bm{b}$ of a given class to its uniform representative $\bm{x}^0\sslash \bm{x}^0$, such that 
\begin{equation} \label{e:dosfis}
\bm{x}^0 = \bm{\Phi}_{\bm{b}}(\bm{x}), \ \ \ \Leftrightarrow \ \ \ \bm{x} \sslash \bm{b} \sim \bm{x}^0 \sslash \bm{x}^0.
\end{equation}
If, when shown on a fixed surround $\bm{b}$, the stimuli $\bm{x}^\alpha$ and $\bm{x}^\beta$ are perceived as different, then they necessarily belong to different classes, and $\bm{\Phi}_{\bm{b}}$ maps them to different uniform representatives. Therefore, for fixed $\bm{b}$, the function $\bm{\Phi}_{\bm{b}}(x)$ must be injective. Since $\bm{x} \sslash \bm{b}$ and $\bm{x}^0 \sslash \bm{x}^0$ belong to the same class, the functions $\bm{\Phi}_{\bm{b}}$ and $\bm{\Phi}_{\bm{b}_\alpha \to \bm{b}_\beta}$ must obey the relation
\begin{equation} \label{eq:phigrandechica}
\bm{\Phi}_{\bm{b}^\alpha \to \bm{b}^\beta} =  \bm{\Phi}_{\bm{b}^\beta}^{-1} \circ \bm{\Phi}_{\bm{b}^\alpha},
\end{equation}
where the symbol $\circ$ represents function composition, so that $\bm{\Phi}_{\bm{b}_\beta}^{-1} \circ \bm{\Phi}_{\bm{b}_\alpha}(\bm{x}_\alpha) \equiv \bm{\Phi}_{\bm{b}_\beta}^{-1}\left[\bm{\Phi}_{\bm{b}_\alpha}(\bm{x}_\alpha) \right]$. The injectivity of $\bm{\Phi}_{\bm{b}}$ guarantees that the inverse $\bm{\Phi}_{\bm{b}}^{-1}$ exists. 

Uniform representatives remain unchanged by $\bm{\Phi}$, that is, $\bm{\Phi}_{\bm{x}}(\bm{x}) = \bm{x}$, for all $\bm{x}$. The uniqueness of uniform representatives implies that all the points along the diagonal correspond to different classes, and that classes must cross the diagonal once and only once.


\subsection{A notion of distance in color space}
\label{sect:esta}

From the above considerations, it follows that any notion of distance between colors must be expressible as a notion of distance between classes of equivalence. That is, distances are objects of the form $d([\bm{x}^\alpha \sslash \bm{b}^\alpha],[\bm{x}^\beta \sslash \bm{b}^\beta])$, where the square brackets $[\cdot]$ represent the class of the enclosed pair. In this section, we establish a mathematical relation between the sought distance and the function $\bm{\Phi}_{\bm{b}}$ defined in the previous section. To simplify the notation, from now on we omit the square brackets, writing $d(\bm{x}^\alpha \sslash \bm{b}^\alpha,\bm{x}^\beta \sslash \bm{b}^\beta)$ to represent $d([\bm{x}^\alpha \sslash \bm{b}^\alpha],[\bm{x}^\beta \sslash \bm{b}^\beta])$. Moreover, when the pairs are uniform representatives, we write $d(\bm{x}^\alpha,\bm{x}^\beta)$ to represent $d([\bm{x}^\alpha \sslash \bm{x}^\alpha],[\bm{x}^\beta \sslash \bm{x}^\beta])$. Therefore, although each argument of the distance function may appear to be a pair, or even a single chromaticity, readers should be aware that arguments are always classes. In other words,
\begin{eqnarray} 
    d(\bm{x}^\alpha \sslash \bm{b}^\alpha, \bm{x}^\beta \sslash \bm{b}^\beta) &:=& d([\bm{x}^\alpha \sslash \bm{b}^\alpha], [\bm{x}^\beta \sslash \bm{b}^\beta]) \nonumber \\
     d(\bm{x}^\alpha, \bm{x}^\beta) &:=&  d([ \bm{x}^\alpha \sslash \bm{x}^\alpha], \ [\bm{x}^\beta \sslash \bm{x}^\beta]). \nonumber
\end{eqnarray}

If distances are properties of whole classes, then the distance between non-uniform stimuli is equal to the distance between the corresponding representatives,
\begin{eqnarray} 
d(\bm{x}^\alpha \sslash \bm{b}^\alpha, \bm{x}^\beta \sslash \bm{b}^\beta) &=& d\left(\bm{\Phi}_{\bm{b}^\alpha}(\bm{x^\alpha})\sslash \bm{\Phi}_{\bm{b}^\alpha}(\bm{x^\alpha}), \bm{\Phi}_{\bm{b}^\beta}(\bm{x^\beta})\sslash \bm{\Phi}_{\bm{b}^\beta}(\bm{x^\beta}) \right) \nonumber \\
&=& d \left(\bm{\Phi}_{\bm{b}^\alpha}(\bm{x}^\alpha), \bm{\Phi}_{\bm{b}^\beta}(\bm{x^\beta})\right). \label{e:igualdad}
\end{eqnarray}
Hence, if the distance between uniform representatives is known (along the white diagonal in Fig.~\ref{f0001}), to calculate the distance between two pairs that do not both lie along the diagonal, we must first slide them through their respective classes of equivalence until they both hit the diagonal (in general, on different places), and then use the definition of distance for uniform representatives.

\begin{figure}
    \centering
    \includegraphics[width=\textwidth,keepaspectratio]{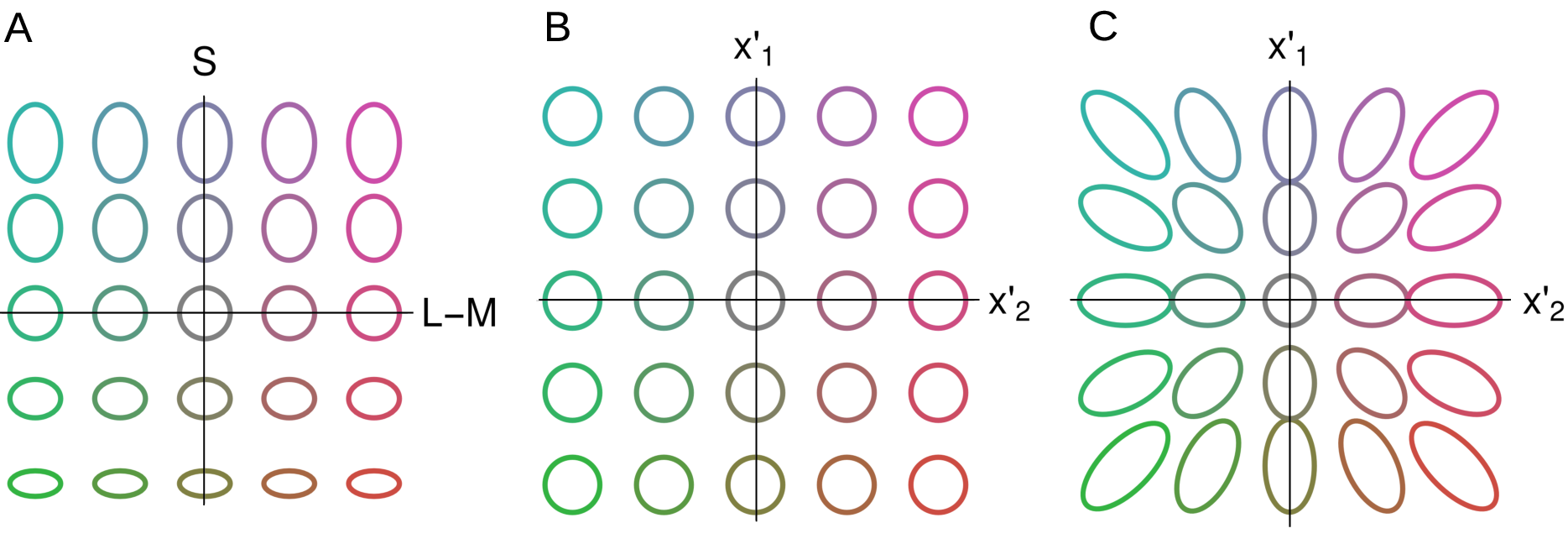}
    \caption{Discrimination Ellipses when stimulus coincides with surrounds, measured in (A):$(S,L-M)$  coordinates, as reported by \cite{Krauskopf1992}, and (B) perceptual coordinates, by definition of the perceptual coordinates. If the surround has a chromaticity that differs from that of the stimuli and is at origin, when including the induction effect produced by a radial and isotropic $\bm{\Phi}_{\bm{b}}(\bm{x})$, the ellipses take a radial form in perceptual coordinates (C), an effect that is conceptually similar to the one reported by \cite{Krauskopf1992}. }
    \label{fig:m}
\end{figure}

\begin{figure}[ht]
\centering
\includegraphics[scale=0.7]{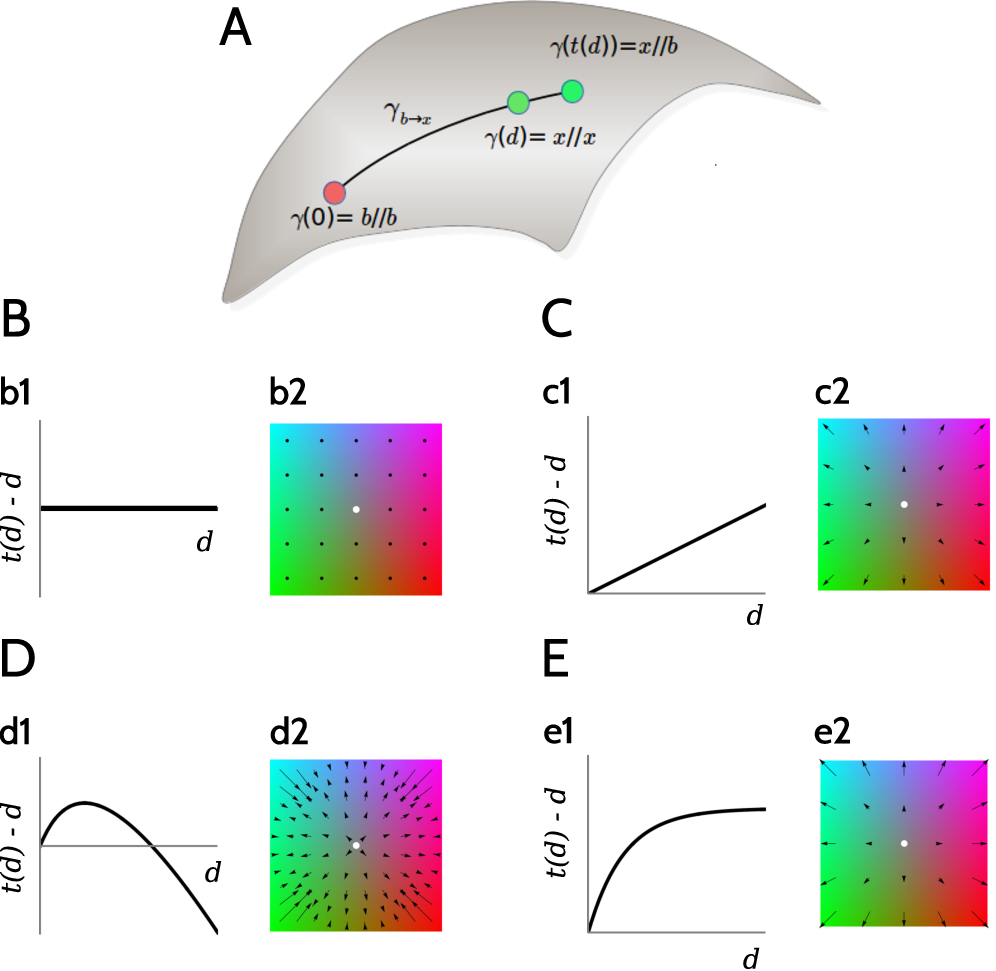}\caption{{\bf Radially symmetric induction}. 
{\sl A}: Hypothesis $4$ and $5$ state that when chromaticity $\bm{x}$ is 
surrounded by chromaticity $\bm{b}$, the perceived sensation is chromatically equal 
to that of a uniform representative that lies along the geodesic 
$\bm{\gamma}_{\bm{b}\to \bm{x}}$, displaced from $\bm{x}$ in an amount $t(d) - 
d$. The space ${\cal P}$ contains all uniform representatives. In this example, the 
surround exerts a repulsive effect, since the 
$\bm{\gamma}_{\bm{b}\to\bm{x}}\left(t(d)\right)$ is longer than 
$\bm{\gamma}_{\bm{b}\to\bm{x}}(d)$. {\sl B-E}: Four different examples of the shifts 
$t(d) - d$, corresponding to the classes of equivalence of Fig.~\ref{f0001}. {\sl B}: 
$t(d) = d$ (b1), and the vector field centered at the surround (white disk) vanishes 
in all the points of color space (b2).  {\sl C}: $t(d) \propto d$, with a proportionality 
factor different from unity. The vector field is linear. For $\bm{x} = \bm{b}$ the 
surround does not alter the perceived stimulus, but otherwise, the effect is radial, 
repulsive, and proportional to the distance between $\bm{x}$ and $\bm{b}$. {\sl D}: 
$t(d)  \propto \ln (1 + d / \lambda)$, for some characteristic distance $\lambda$. The 
effect o f the surround is initially repulsive, vanishes at $d = \lambda$, and then 
reverts to attractive. In {\sl E}, $t(d)-d  \propto [1 - \exp(-d /\lambda)]$, so the 
displacement is always repulsive, and tends to a constant value for large distances. 
\label{f0002}} 
\end{figure}

Below we list the hypothesis under which we construct the geometry of color space.
\begin{enumerate}
    \item \emph{The manifold of percepts is Riemannian}, so that the distance function $d$ can be written in terms of a metric tensor $J$. This hypothesis is implicit in Eqs.~\ref{e:infdistan}-\ref{e:x2prima}. The Riemannian assumption was first introduced by \cite{Helmholtz1892} and  \cite{Schrodinger1920}, later discussed by \cite{Silberstein1943},  \cite{Stiles1946},  \cite{Resnikoff1974}, and \cite{Fonseca2016, Fonseca2018}, and is supported by the experimental observation that discrimination thresholds conform an ellipse around the reference color \cite{MacAdam1944, Krauskopf1992}, so local distances can be approximated by a quadratic form (Fig.~\ref{fig:m}). 
    
    \item \emph{The metric tensor is decomposable as a direct sum} in the isoluminant coordinates  $x_1 = S$ and $x_2 = L - M$. This hypothesis is implicit in Eqs.~\ref{e:jepsilon}-\ref{e:x2prima}, and yields
    \begin{equation}
        d\ell^2 = J_{11}(x_1) \ \mathrm{d}x_1^2 + J_{22}(x_2) \ \mathrm{d}x_2^2.
    \end{equation}
    The assumption derives from the fact that discrimination ellipses, as shown by \cite{Krauskopf1992}, have principal axes that are parallel to the cardinal axes $\hat{\bm{e}}^1$ and $\hat{\bm{e}}^2$ (Fig.~\ref{fig:m}A). Under this assumption, the space is flat, and a coordinate system exists (the {\sl perceptual coordinates}) in which the distance between uniform representatives is Euclidean (Sect.~\ref{sec:natural}). In this coordinate system, discrimination ellipses are circles, and all have the same size (Fig.~\ref{fig:m}B).
    
    \item \emph{The space of percepts is complete}, so for any pair of points $\bm{x}^\alpha, \bm{x}^\beta$, a geodesic $\gamma_{\bm{x}^\alpha \to \bm{x}^\beta}$ joining them exists such that $d(\bm{x}^\alpha, \bm{x}^\beta) = \mathrm{length}( \gamma_{\bm{x}^\alpha \to \bm{x}^\beta})$. In particular, the separability of the isoluminant plane implies that the lines defined by the cardinal axes $\bm{e}^1$ and $\bm{e}^2$ are geodesics.
\end{enumerate}
One of the central hypotheses of this paper is that, in the perceptual coordinates, the effect of the surround has rotational symmetry.  More precisely, the function $\bm{\Phi}_{\bm{b}}(\bm{x})$ is assumed to comply with two other requirements: 
\begin{enumerate}\setcounter{enumi}{3}
    \item \emph{The radial hypothesis}: If $\bm{x}^\alpha\sslash \bm{b}^\alpha \sim \bm{x}^\beta \sslash \bm{b}^\beta$, and $\bm{x}^\alpha, \bm{b}^\alpha$ and $\bm{b}^\beta$ lie all on the same cardinal axis (either $\hat{\bm{e}}^1$ or $\hat{\bm{e}}^2$), the matched chromaticity $\bm{x}^\beta$ also lies on the same axis. Evidence for this symmetry is discussed in {\sl Experiment III}. So far, this hypothesis was formulated for the cardinal axis of the cone contrast coordinates. To make the statement more general, we observe that the conjecture suggests that the displacement produced by $\bm{\Phi}_{\bm{b}}(\bm{x})$ acts along the line connecting the stimulus and the surround, such that for fixed $\bm{b}$, the vector field of displacements induced by $\bm{\Phi}_{\bm{b}}(\bm{x})$ is radial and centered in $\bm{b}$. Graphically, in the vector fields of  Fig.~\ref{f0002}, arrows are parallel to the line joining $\bm{b}$ and $\bm{x}$. In Riemannian geometries, the line connecting two points is generalized to a geodesic (Fig.~\ref{f0002}A), so the precise formulation of the radial hypothesis reads: For fixed $\bm{b}$ and viewed as a function of $\bm{x}$, the uniform representative $\bm{\Phi}_{\bm{b}}(\bm{x})$ lies along the geodesic $\bm{\gamma}_{\bm{b} \to \bm{x}}$ that starts from $\bm{b}$ and passes through $\bm{x}$. Moreover, if $t$ is an arc-length affine parameter for $\bm{\gamma}_{\bm{b} \to \bm{x}}(t)$, a scalar function $t(\bm{x}, \bm{b})$ exists, such that the uniform representative can be written as $\bm{\Phi}_{\bm{b}}(\bm{x}) = \bm{\gamma}_{\bm{b} \to \bm{x}}[t(\bm{x}, \bm{b})]$.
    
    \item \emph{Isotropy and homogeneity}: Color space is assumed to contain no privileged stimuli or directions, at least, when dealing with points that are far from the borders of the gamut (stimuli that are maximally saturated). Evidence for this hypothesis is provided by {\sl Experiments II} and {\sl III}. The core assumption is that the perceptual shift produced by a surround $\bm{b}$ on a stimulus $\bm{s}$ only depends on the distance $d(\bm{b}, \bm{x})$, that is, $t(\bm{x},\bm{b}) = t[d(\bm{x}, \bm{b})]$. The perceptual coordinates are defined so as to ensure that equi-distant classes cross the diagonal in equi-distant points. Yet, from the definition of perceptual coordinates alone, there is no restriction on the shape of classes. The isotropy and homogeneity hypothesis implies that, when viewed in the perceptual coordinates, all the classes have the same shape, and only differ from one another in a rigid translation, as in all the examples of Fig.~\ref{f0001}.  
\end{enumerate}

In the perceptual coordinates, the metric tensor reduces to the unit matrix, so all geodesics become straight lines, along which components can be summed and multiplied. In particular, the separability of the metric tensor (hypothesis 2) implies that the lines along the cardinal axes $\hat{\bm{e}}^1$ and $\hat{\bm{e}}^2$ are geodesics. In the perceptual coordinates, the mapping $\bm{\Phi}_{\bm{b}}(\bm{x})$ can be written as
\begin{equation} \label{eq:dispnat}
\bm{\Phi}_{\bm{b}}(\bm{x}) = \bm{\gamma}_{\bm{b}\to \bm{x}}\left(t\left(d(\bm{x}, \bm{b})\right)\right) = \bm{b} + t \left(d(\bm{b}, \bm{x})\right) \ \hat{\bm{u}}, \ \ \ \mathrm{with} \ \hat{\bm{u}} = \frac{\bm{x} - \bm{b}}{d(\bm{b}, \bm{x})}.
\end{equation}
That is, in these coordinates, the color shift induced by the surround is radial, it is centered at the surround $\bm{b}$ and is of magnitude $t\left(d(\bm{x}, \bm{b})\right)$ along the direction $\hat{\bm{u}}$ that connects $\bm{b}$ and $\bm{x}$. Figure \ref{fig:m}~C displays example discrimination ellipses for a fixed surround when $\bm{\phi_b}$ satisfies hypotheses 4 and 5.

If the surround exerts no influence (Fig.~\ref{f0001}A) then $\bm{\Phi}_{\bm{b}}(\bm{x}) = \bm{x}$, which necessarily implies that $t(d) = d$.  

The effect of the surround is taken to be {\sl repulsive} if $t(d) > d$ (the surround repels the stimuli, so that the uniform representative of a given stimulus is further away from the surround than the original stimulus), and {\sl attractive} otherwise, that is, if $t(d) < d$. 


\subsection{Experiment I: Discrimination thresholds for $B = T$ }
\label{s:exp1}

{\sl Experiment I} is used to find the perceptual coordinates of each observer, for which the knowledge of the metric tensor $J(\bm{x})$ is required (Eqs.~\ref{e:x1prima} and \ref{e:x2prima}). We work with fixed luminosity, that is, $L + M = \mathrm{const}.$ \cite{Derrington1984}. 
 
 \cite{Krauskopf1992} established that in the space $(x_1, x_2) = (S,L-M)$ defined by the cone contrasts, discrimination thresholds are described by diagonal quadratic forms (hypothesis 2 above). Moreover, the elongation of the ellipses along the $\hat{\bm{e}}^1$ direction varied approximately linearly with $x_1$, and bared no significant dependence on $x_2$. The elongation along the $\hat{\bm{e}}^2$ direction was shown to be approximately constant. These results  imply that it suffices to sample the thresholds around colors $\bm{x}$ that lie along the cardinal axes, testing displaced colors $\bm{x} + \varepsilon_I \hat{\bm{e}}$ that also lie along the same axis. The colors $\bm{x}$ tested here are displayed in Fig.~\ref{f1}{\sl C}.

We now deduce how the diagonal terms $J_{ii}$ are obtained from the measured discrimination thresholds. For each uniform representative $\bm{x} \sslash \bm{x}$ sampled along the $i$-th coordinate axis ($i \ \in \ \{1, 2\}$), we determine the minimal displacement $\varepsilon_I({\bm{x}}, \hat{\bm{e}}^i)$ along the same direction $\hat{\bm{e}}^i$, so that $\bm{x} + \varepsilon_I({\bm{x}}, \hat{\bm{e}}^i) \hat{\bm{e}}^i \sslash \bm{x}$ be first distinguishable from ${\bm x}\sslash{\bm x}$. The sub-index ``$I$'' in $\varepsilon_I$ indicates a threshold obtained with {\sl Experiment I} (a different threshold is defined in {\sl Experiment II}). 

Defining the unit of distance in color space as that corresponding to the just noticeable difference (Sect.~\ref{s:methods:discrimination}), and making use of the assumption that distances derive from a diagonal metric tensor $J$,
\begin{align}
    1 &= d\left(\bm{x} \sslash \bm{x}, \bm{x} + \varepsilon_I({\bm{x}}, \hat{\bm{e}}^i)\hat{\bm{e}}^i \sslash \bm{x}\right) \nonumber \\
      &= d\left( \bm{x}, \bm{\Phi}_{\bm{x}} \left(\bm{x} + \varepsilon_I({\bm{x}}, \hat{\bm{e}}^i) \hat{\bm{e}}^i\right) )\right) \nonumber   \\
      &= \mathrm{Length \ of \ the \ geodesic} \ \bm{\gamma}\left( t \left( d\left(\bm{x}, \bm{x} + \varepsilon_I({\bm{x}}, \hat{\bm{e}}^i)\hat{\bm{e}_1}   \right) \right) \right) \nonumber \\
      &= \vert t\left( d\left( \bm{x}, \bm{x} + \varepsilon_I({\bm{x}}, \hat{\bm{e}}^i) \right) \right)\vert \nonumber \\
      &   \approx \vert  t'(0) \sqrt{ J_{ii}(\bm{x})}  \varepsilon_I({\bm{x}}, \hat{\bm{e}}^i) \vert \label{e:teprima}
\end{align}
Two factors determine the length $\varepsilon_I(\bm{x}, \hat{\bm{e}}^i)$ corresponding to the just noticeable difference: The metric $J$, and the derivative $t'(0)$. The metric defines how distances are quantified in each point of color space and along each direction, and appears in any Riemmanian space. The derivative is a special ingredient that appears in the case of {\sl Experiment I}. Note that the metric is evaluated on the color $\bm{x}$, so $J(\bm{x})$ alone does not contain information about the sliding operation that was required to take $\bm{x} + \varepsilon_I({\bm{x}}, \hat{\bm{e}}^i)\hat{\bm{e}}^i$ to its uniform representative $\bm{\Phi}_{\bm{x}} \left(\bm{x} + \varepsilon_I({\bm{x}}, \hat{\bm{e}}^i)\hat{\bm{e}}^i\right) \sslash \bm{\Phi}_{\bm{x}} \left(\bm{x} + \varepsilon_I({\bm{x}}, \hat{\bm{e}}^i)\hat{\bm{e}}^i\right)$. The derivative is precisely the factor that provides that information. Its numerical value may depend on experimental conditions, as the size and geometry of the stimuli \cite{Nagy1993,Kellner2016}, the stimulation time window \cite{Siegel1965}, etc
. 
If the surround exerts no influence (horizontal classes in Fig.~\ref{f0001}A), then the derivative is equal to unity, since $\bm{\Phi}_{\bm{b}}(\bm{x}) = \bm{x}$ and $t(d) = d$. If the surround exerts a repulsive effect, the derivative is larger than unity, since for small distances $t(d) \approx t'(0) d$. This case is illustrated in Fig.~\ref{f0001}, where the contour lines  have positive slope when crossing the diagonal. Repulsive surrounds increase the distance, or equivalently, to reach the same perceptual distance, a smaller threshold suffices. An attractive surround, instead, corresponds to $t'(0) < 1$.

\begin{figure}[ht]
\centering
\includegraphics[scale = 0.85, keepaspectratio]{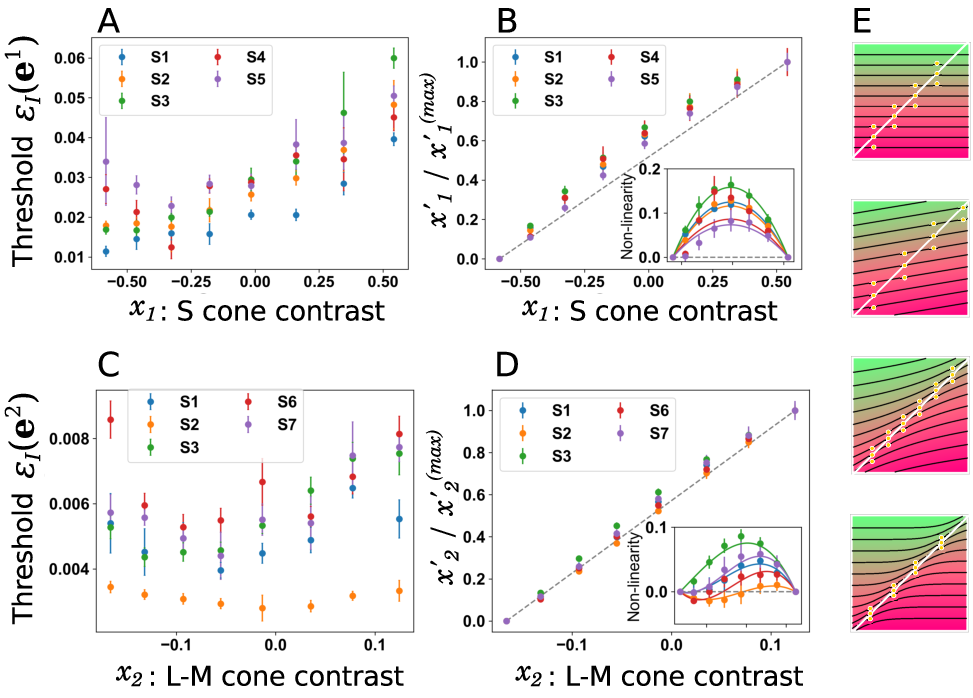}
\caption{{\bf Discrimination thresholds when the target and the surround chromaticities coincide}.  {\sl A} and {\sl C}:  Discrimination thresholds for the $x_1$ ({\sl A}) and $x_2$ ({\sl C}) cone contrast coordinates.  Different observers displayed in different colors. {\sl B} and {\sl D}: perceptual coordinates $x_1'/{x_1'}^{\mathrm{max}}$ ({\sl B}) and $x_2'/{x_2'}^{\mathrm{max}}$ ({\sl D}) as a function of the cone contrasts. The normalizing factors ${x_i'}^{\mathrm{max}}$ are the maximal perceptual coordinate obtained for each subject, and were used to scale the data in order to compare different observers, which would otherwise produce perceptual coordinates spanning intervals of different lengths. Insets: Deviations from the linear mapping. Each data point is obtained from the fit of Eq.~\ref{eq:asimetria}, and error bars are the expected errors of the fit. Parameters of the optimal fits are given in Table~\ref{t1}. {\sl E}: The measured thresholds represent the vertical displacement between a pair $\bm{x} \sslash \bm{x}$ on the diagonal (yellow dot), and another pair sitting right above, or just below, on a class of equivalence that is at perceptual distance $1$ from that of $\bm{x} \sslash \bm{x}$. 
\label{f3} } 
\end{figure}

Solving Eq.~\ref{e:teprima} for $J_{ii}$,
\begin{equation}
\label{eq: exp1.1}
    J_{ii}(\bm{x}) = \left[\frac{1}{t'(0) \ \varepsilon_I(\bm{x}, \hat{\bm{e}}^i)}\right]^{2}.
\end{equation}
This equation allows us to find the diagonal terms $J_{ii}$ from the thresholds $\varepsilon_I(\bm{x}, \hat{\bm{e}}^i)$, up to a multiplicative factor $t'(0)$. Figures~\ref{f3}A and C show the measured thresholds.
Along the $x_1$ axis, thresholds increased roughly linearly with $x_1$, with some subject-to-subject variability. Thresholds varied across subjects up to a factor of 3. Along the $x_2$ axis, thresholds showed a non monotonic behavior, with a minimum around $x_2 = 0$, which corresponds to the reference gray. Although there was a certain subject-to-subject variability, all observers showed evidence of the minimum. For each 
subject, the modulation of thresholds along this axis was significantly smaller than along the $S$ axis, with the maximal and minimal threshold of each observer differing in less than $50\%$ of the minimal threshold. Hence, confirming the result of \cite{Krauskopf1992}, thresholds along the $\hat{\bm{e}}^1$ direction vary more pronouncedly than along the $\hat{\bm{e}}^2$ direction. Yet, our data reveal that they do not remain strictly constant along the $\hat{\bm{e}}^2$ directions, since the mild non-monotonic behavior was found to be significant.

The threshold $\varepsilon_I(\bm{x}, \hat{\bm{e}}^i)$ was defined as the change in chromaticity required for a stimulus to be discriminated from its surround  in 62.5\% of the trials (Sect.~\ref{s:methods:discrimination}). In Fig.~\ref{f3}{\sl E}, this increment is the vertical displacement between a pair $\bm{x} \sslash \bm{x}$ on the diagonal, and a point $\bm{x} + \varepsilon_I(\bm{x}, \hat{\bm{e}}^i)\hat{\bm{e}}^i \sslash \bm{x}$ sitting right above (or below) the former, on the equivalence class at distance $1$ from that of $\bm{x} \sslash \bm{x}$.  If Fig.~\ref{f3}{\sl E} were depicted in cone  coordinates or in any other color space that had not been chosen to reflect perceptual distances, different triplets of yellow dots along the diagonal would appear to span different vertical heights, since the separation just-noticeable-different classes (classes at distance $1$) can be arbitrary. Finding the perceptual coordinates is equivalent to finding a representational system in which the vertical span of all triplets remain constant along the diagonal. In these coordinates, the classes intersect the diagonal at equi-distant intervals, as in Figs.~\ref{f0001} and \ref{f3}{\sl E}. 

To define the perceptual coordinates along the axes $\hat{\bm{e}}^1$ and $\hat{\bm{e}}^2$, the square root of the diagonal elements of the metric have to be integrated (Eq.~\ref{e:distancia}). To this aim, an analytic expression of $J^{\sfrac{1}{2}}(\bm{x})$ is needed. We proposed a polynomial function 
\begin{equation} \label{e:poly}
|t'(0)| \ \sqrt{J_{ii}(x_i)}  = \frac{1}{\varepsilon_I(\bm{x}, \hat{\bm{e}}^i)} = \sum_{j = 0}^{n} \alpha_j \ x_i^{j},
\end{equation}
and fitted the coefficients $\alpha_j$ to the data. The order $n$ of the polynomial was chosen as the lowest that still accounted for the data with $p$-values above $0.01$. Along the $\hat{\bm{e}}^1$ axis, a straight line ($n = 1$) suffices, whereas the $\hat{\bm{e}}^2$ axis requires to go up to a quadratic expression ($n = 2$). Table~\ref{t1} of the Appendix contains the fitted parameters.

Along the $\hat{\bm{e}}^1$ direction, the variability of the coefficients fitted for different observers indicated inter-individual differences, since a single set of coefficients $\alpha_j$ could not account for the metric tensor of different subjects. The $p$-value for the hypothesis that a single $\alpha_0$ could be used for the $5$ subjects was $10^{-8}$, and for a single $\alpha_1$ was $6 \cdot 10^{-3}$. Along the $\hat{\bm{e}}^2$ direction, the individual differences were significant in the constant ($p$-value below $10^{-10}$) and linear coefficients ($p$-value $2 \ 10^{-7}$), but not in the quadratic ones ($p$-value = 0.68).

Once an analytic expression has been obtained for the diagonal elements of the metric, the perceptual coordinates along the cardinal axes can be calculated by integration (Eqs.~\ref{e:x1prima} and \ref{e:x2prima}), except for the yet unknown factor $|t'(0)|$. In Figs.~\ref{f3}{\sl B} and {\sl D}, the normalized perceptual coordinates $x_1'$ and $x_2'$ are shown as a function of the corresponding cone contrasts $x_1$ and $x_2$. The insets display the deviation from a linear mapping, together with the quadratic or cubic analytical expressions obtained by integrating Eq.~\ref{e:poly} (same parameters as in Table~\ref{t1}). Importantly for what follows, in the perceptual coordinates, the distance between two colors $\bm{x}'$ and $\bm{y}'$ is calculated with the Euclidean formula. If the two colors lie along the cardinal axis $\hat{\bm{e}}^i$, then $d(x'\hat{\bm{e}}^i, y'\hat{\bm{e}}^i) = |x'_i - y'_i|$.


\subsection{Experiment II: Discrimination Thresholds for $B \ne T$} 
\label{s:exp2}

{\sl Experiment II} involved the same discrimination task as {\sl Experiment I}, but with a surround $\bm{b}$ that was different from the tested stimuli. Since the discrimination threshold depends  on the surround, we use the notation $\varepsilon_{II}(\bm{x}, \bm{b}, \hat{\bm{e}}^i)$. {\sl Experiment II} reduces to {\sl Experiment I} when $\bm{b} = \bm{x}$, that is, $\varepsilon_{II}(\bm{x}, \bm{x}, \hat{\bm{e}}^i) \equiv \varepsilon_I(\bm{x}, \hat{\bm{e}}^i)$. 

In {\sl Experiment II},
\begin{align}
    1 &=  d \left( \bm{x} \sslash \bm{b}, \bm{x} + \varepsilon_{II}(\bm{x},  \bm{b}, \hat{\bm{e}}^i) \hat{\bm{e}}^i \sslash \bm{b} \right) \nonumber \\
      &=  d\left(\bm{\Phi}_{\bm{b}} (\bm{x}), \bm{\Phi}_{\bm{b}}  \left( \bm{x} + \varepsilon_{II}(\bm{x}, \bm{b}, \hat{\bm{e}}^i) \hat{\bm{e}}^i \right) \right), \label{e:exp2}
\end{align}
where the second line derives from the hypothesis that distances between two pairs remain invariant if any of the pairs is replaced by another member of its own class, in particular, the uniform representative. Since $\bm{b}$, $\bm{x}$ and $\bm{x} + \varepsilon_{II}(\bm{x}, \bm{b}, \hat{\bm{e}}^i) \hat{\bm{e}}^i$ lie all three on the same cardinal axis,
\[
    d\left( \bm{\Phi}_{\bm{b}} (\bm{x}), \bm{\Phi}_{\bm{b}} \left( \bm{x} + \varepsilon_{II}(\bm{x}, \bm{b}, \hat{\bm{e}}^i) \hat{\bm{e}}^i \right) \right) = \left\vert  d\left( \bm{\Phi}_{\bm{b}}(\bm{x} + \varepsilon(\bm{x}, \bm{b}, \hat{\bm{e}}^i)  \hat{\bm{e}}^i), \bm{b}\right)- d\left( \bm{\Phi}_{\bm{b}}(\bm{x}), \bm{b} \right) \right\vert.
\]
Replacing this result in Eq.~\ref{e:exp2},      
\begin{align}
    1 &= \left\vert t\left( d\left( \bm{x} + \varepsilon_{II}(\bm{x}, \bm{b}, \hat{\bm{e}}^i)\hat{\bm{e}}^i, \bm{b}\right) \right) - t\left( d(\bm{x}, \bm{b})\right) \right\vert \nonumber \\
    & \approx  \left\vert t'\left( d(\bm{x}, \bm{b})\right) \sqrt{J_{ii}(\bm{x})} \varepsilon_{II}(\bm{x}, \bm{b}, \hat{\bm{e}}^i) \right\vert 
\end{align} 
Since $J_{ii}$ is known from {\sl Experiment I}, we can use Eq.~\ref{eq: exp1.1} to get
\begin{equation} \label{e:e2epsilon2}
  \varepsilon_{II}(\bm{x}, \bm{b}, \hat{\bm{e}}^i) =  \varepsilon_I(\bm{x}, \hat{\bm{e}}^i) \left|\frac{t'(0)}{t'[d(\bm{x}, \bm{b})]} \right|. 
\end{equation}
In the perceptual coordinates, $\varepsilon'_I(\bm{x}, \hat{\bm{e}}^i) =1$, by definition. Therefore, if $\varepsilon_{II}(\bm{x}, \bm{b}, \hat{\bm{e}}^i)$ is written in the perceptual coordinates and the isotropy and homogeneity hypotheses hold, Eq.~\ref{e:e2epsilon2} implies that the thresholds in {\sl Experiment II} depend only on the distance between stimulus and surround, irrespective of the specific surround or cardinal axis. In the perceptual coordinates, distances between stimuli belonging to the same cardinal axis are simply equal to $\vert x'_i - b'_i \vert $, so $\varepsilon_{II}(\bm{x}, \bm{b},\hat{\bm{e}}^i)$ depends on its arguments only through the combination $\vert x'_i - b'_i\vert$.

When the surround coincides with the stimulus, we get $\varepsilon_I = \varepsilon_{II}$. As the surround $\bm{b}$ is moved away from the stimulus $\bm{x}$, the distance $d(\bm{x}, \bm{b})$ increases. The threshold $\varepsilon_{II}$ may then either increase or decrease, depending on whether the absolute value of the slope of $t(d)$ is larger or smaller than  $t'(0)$. Therefore, by measuring the thresholds $\varepsilon_{II}$ for different surrounds, the derivative of $t(d)$ is revealed. Yet, this reasoning is only valid if the \emph{isotropy and homogeneity hypothesis} proposed above (number 5 in Sect.~\ref{sect:esta}) indeed holds, namely, the assumption that the perceptual shift induced by the surround only depends on the distance $d(\bm{x}, \bm{b})$. Therefore, before characterizing the shape of $t(d)$, we first use {\sl Experiment II} to assess the validity of this hypothesis. To do so, we demonstrate that, in the perceptual coordinates, the dependence of the thresholds $\varepsilon_{II}(\bm{x}, \bm{b}, \hat{\bm{e}}^i)$ with $\bm{b}$ and with $\bm{x}$ can be entirely written in terms of the distance $\vert x'_i - b'_i \vert$.

\begin{figure}[ht]
\centering
\includegraphics[scale=0.11]{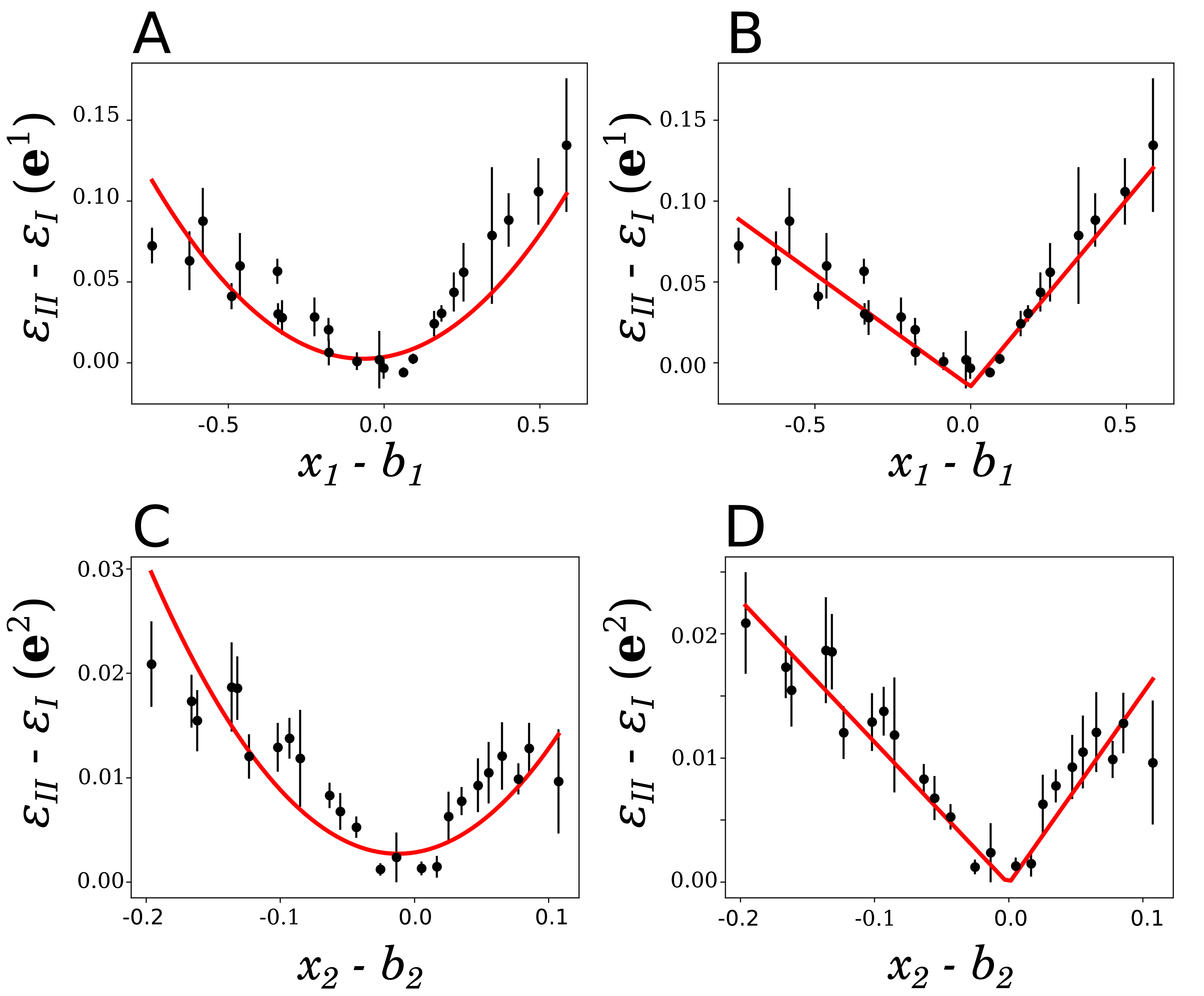}
\caption{{\bf Performance of models 1 and 2 in describing measured thresholds $\varepsilon_{II}(\bm{x}, \bm{b}, \hat{\bm{e}}^i)$.} Thresholds were measured for observer S2, and are shown as a function of $x_i - b_i$. Data points are obtained from the fit of Eq.~\ref{eq:asimetria}, and error bars are the expected error of the fit. Red line: fitted model. \emph{A} and \emph{C}: Model 1 (Eq.~\ref{e:exp2mod1}). \emph{B} and \emph{D}: Model 2 (Eq.~\ref{e:exp2mod2}). \emph{A} and \emph{B}: Discrimination thresholds measured along $\hat{\bm{e}}^1$.  \emph{C} and \emph{D}: Along $\hat{\bm{e}}^2$. }
\label{f4} 
\end{figure}

The first step is to describe the dependence of the thresholds on the surround in the cone contrast coordinates. In Fig.~\ref{f4}, we see the variation of the thresholds from those obtained in {\sl Experiment I} of a given subject as a function of the difference $x_i - b_i$. As reported by \cite{Krauskopf1992},
the thresholds are minimal for $\bm{b} = \bm{x}$, and increase as the surround 
differs from the stimulus. This non-monotonic behavior refutes the hypothesis that 
classes be linear functions of the stimulus, as proposed by \cite{Resnikoff1974}. It 
then becomes important to characterize the variation.  In {\sl Experiment II}, the 
surround is always relatively close to the stimulus, so an expansion of 
$\varepsilon_{II}(\bm{x}, \bm{b}, \hat{\bm{e}}^i)$ around $\bm{b} = \bm{x}$ can be 
used to describe the measured thresholds.

If, as assumed in this paper, the function $t$ depends on the coordinates through the distance $d$, the first-order of  the Taylor expansion of $t'$ must include the term $\vert x_i - b_i \vert$. Alternatively, if {\sl Hypotheses 4} and 5 do not hold, thresholds would be expected to vary smoothly with the coordinates, in which case, a polynomial would provide a reasonable description of the dependence. We therefore compare two models containing the same number of parameters:
\begin{eqnarray}
\mathrm{Model \ }1: \ \ \ \varepsilon_{II}(\bm{x}, \bm{b}, \hat{\bm{e}}^i) - \varepsilon_I(\bm{b}, \hat{\bm{e}}^i) &\approx&  \gamma_0 + \gamma_1 (x_i - b_i) + \gamma_2 (x_i - b_i)^2 \label{e:exp2mod1} \\
\mathrm{Model \ }2: \ \ \ \varepsilon_{II}(\bm{x}, \bm{b}, \hat{\bm{e}}^i) - \varepsilon_I(\bm{b}, \hat{\bm{e}}^i) &\approx& \gamma_0 + \gamma_1 (x_i - b_i) + \gamma_2 |x_i - b_i|  \label{e:exp2mod2} 
\end{eqnarray}
The first model assumes that $\varepsilon_{II}(\bm{x}, \bm{b}, \hat{\bm{e}}^i)$ has a continuous derivative at $b_i = x_i$, and is able to describe the quadratic departure from linearity. The second model allows for the possibility of a discontinuous derivative, and for the ascending and the descending linear portions to have different slopes. It cannot, however, describe quadratic effects. 

In Fig.~\ref{f4}, we compare the performance of the two proposals in fitting the measured thresholds.
The fitted coefficients $\gamma_0$, $\gamma_1$ and $\gamma_2$ are reported in Tables \ref{t:e2ejeS} and \ref{t:e2ejeLM} of the Appendix. The constant term $\gamma_0$ is of the order of the experimental error of the measurements, confirming that when the stimulus and the surround coincide, $\varepsilon_{II}$ indeed reduces to $\varepsilon_{I}$.

Each fit produces a $\chi^2$ value quantifying the goodness of the fit for each subject and axis, and although there are small differences among conditions, the mean $\chi^2$-value obtained for Model 2 (averaged across subjects and axes) is half the value obtained for Model 1. Accordingly, the mean $p$-value obtained for the hypothesis that the data be generated with Model 2 is twice as large as with Model 1. These results imply that the data is better explained by Model 2, and a discontinuous derivative is to be expected at $\bm{b} = \bm{x}$. Moreover, the fact that $\gamma_1$ is typically significantly different from zero indicates that the ascending and the descending linear portions of Model 2 have different slopes.

\begin{figure}[ht]
\centering
\includegraphics[width=\textwidth, keepaspectratio]{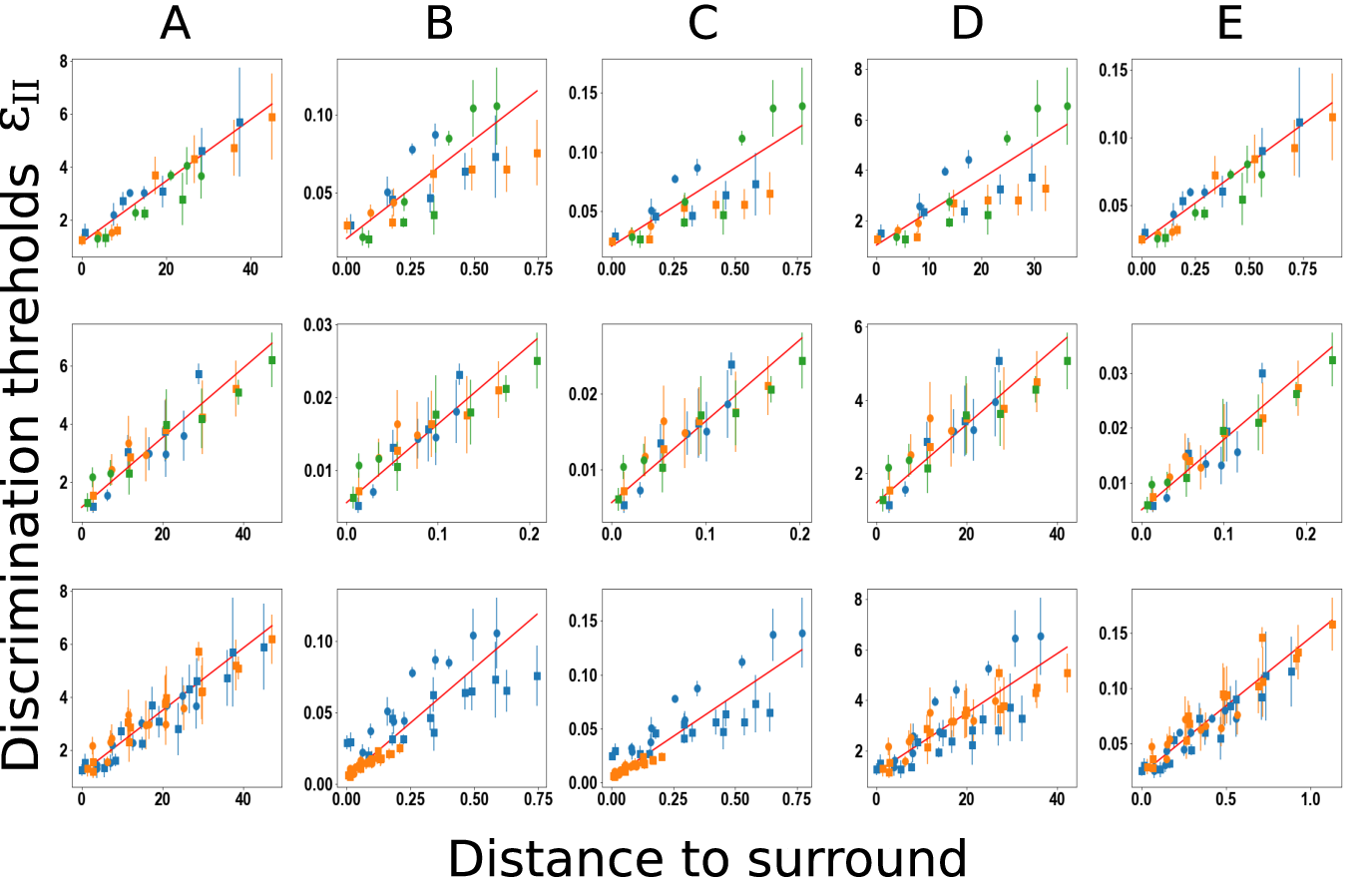}
\caption{{\bf Assessment of the validity of hypothesis 5}. Thresholds measured in {\sl Experiment II} for subject S1 as a function of the distance between the surround and the stimulus. Circles: $b_i > x_i$. Squares: $b_i < x_i$. Each column represents a different choice of the system of coordinates in which thresholds, stimuli and surround are represented. {\sl A}: perceptual coordinates defined with the data of Experiment I. Columns {\sl B, C, D}: Other coordinates employed in the literature (see text), requiring no fitted parameters. Column {\sl E}: Optimal coordinate system defined with a single fitted parameter. Top row: $\bm{b}$ and $\bm{x}$ lie along axis $\hat{\bm{e}}^1$. Green: $\bm{b} = (x_1, x_2) =  (0.16,0)$, blue: $\bm{b} = (0,0)$, orange: $\bm{b} = (-0.24,0)$.  Middle row: axis $\hat{\bm{e}}^2$. Green: $\bm{b} = (0, 0)$, blue: $\bm{b} = (0,-0.03)$, orange: $\bm{b} = (0,0.03)$. Bottom: both axes together. Blue data points: $\hat{\bm{e}}^1$. Orange: $\hat{\bm{e}}^2$.
}
\label{f5} 
\end{figure}

To determine whether the hypothesis of homogeneity and isotropy is justified, we now transform $\bm{x}, \bm{b}$ and $\varepsilon_{II}$ to the perceptual coordinates, using Eqs.~\ref{e:x1prima} and \ref{e:x2prima} and the metric tensor $J_{ii}$ obtained with {\sl Experiment I}. We emphasize that no data of {\sl Experiment II} is used to fit the parameters of the transformation. Although we still lack the multiplicative constant $|t'(0)|$, we can nevertheless assess whether, in these coordinates, $\varepsilon'_{II}(\bm{x}', \bm{b}', \hat{\bm{e}}^i)$ indeed depends only on the difference $|x'_i - b'_i|$. If it does, the transformation should suffice to eliminate the asymmetry in the slopes of the descending and ascending portions of Model 2. Equivalently, when $\varepsilon'_{II}$ (measured for a single subject with different stimuli $\bm{x}$, surrounds $\bm{b}$ and axes $\hat{\bm{e}}^i$) is plotted as a function of $|x'_i - b'_i|$, a single straight line should be seen. This plot is displayed in column A of Fig.~\ref{f5}, for surrounds varying along the axis $\hat{\bm{e}}^1$ (top), $\hat{\bm{e}}^2$ (middle) and both axes together (bottom). 

For comparison, we also show the same data points represented in other coordinate systems, to test whether the linear relation between $\varepsilon'_{II}(\bm{x}', \bm{b}', \hat{\bm{e}}^i)$ with $|x'_i - b'_i|$ indeed becomes more evident in the  perceptual coordinates than in other coordinate systems. In column {\sl B}, the data are plotted in cone contrast coordinates. Clearly, the points obtained for $b_i > x_i$ (circles) define a different slope from that for $b_i < x_i$ (squares). Moreover, the slopes along the axes $\hat{\bm{e}}^1$ and $\hat{\bm{e}}^2$ are markedly different (bottom row), and so are the total ranges of the data. As a consequence, the amount of dispersion is larger in the plots of column {\sl B} than in column {\sl A}. The $\chi^2$-values obtained from linear fits in column B are more than three times larger than those in column 1, meaning that the data are more in line with the \emph{Homogeneity and isotropy hypothesis} when plotted in the perceptual coordinates than in the cone contrasts.

In the cone contrast coordinate system, the origin $\bm{x} = \bm{0}$ is gray by convention. Columns {\sl C} and {\sl D} evaluate the performance of two additional coordinate systems, in which cone contrast is determined with respect to the chromatic surround in which the discrimination task was performed. Specifically, if the supra-index $\mathrm{cc}$ represents cone contrasts, in column {\sl C}, the coordinates of both the stimulus and the surround are defined by the relation $x_i^{\mathrm{new}}= x_i^{\mathrm{cc}} / (b_i^{\mathrm{cc}} + 1)$, so that changes in stimuli are represented by the relative contrast to the surround. 

If $\varepsilon_{II}$ depended only on the ratio $x_i / b_i$,  Weber's law \cite{Wyszecki2000} would hold. In column {\sl D}, the transformation is $x_i^{\mathrm{new}}= x_i^{\mathrm{cc}} / \varepsilon_I(\bm{b}, \hat{\bm{e}}^i)^{\mathrm{cc}}$, so that the threshold of the surround always corresponds to unity. If $\varepsilon_{II}$ depended only on the ratio $\varepsilon_{I}(\bm{x}, \hat{e}^i) / \varepsilon_I(\bm{b}, \hat{\bm{e}}^i)$, a modified version of Weber's law, formulated in terms of thresholds, would govern discriminability. The resulting average $\chi^2$ values represent a three-fold (column {\sl C}) and a two-fold (column {\sl D}) increase with respect to the first column. Thus, again, the perceptual coordinates describe better the linear relation. 

While the first four models assessed coordinate systems that contained no free parameters, the last column was constructed by searching for the value of a free coefficient $\alpha$, obtained from a fit to the data, that produced the mapping $x_i^{\mathrm{new}} = x_i^{\mathrm{cc}} + \frac{\alpha}{2} (x_i^{\mathrm{cc}})^2$ with minimal $\chi^2$-value. The improvement, however, was only marginal, with a $\chi^2$-value that was only 6\% smaller than that for the first coordinate system. The perceptual coordinates, hence, achieve almost the same performance as the ones of the last model without
parameters determined from the data of {\sl Experiment II}. 

In Fig.~\ref{f3}{\sl E}, discrimination thresholds are represented as the vertical distance between yellow dots. We could add additional dots to the figure, thereby extending the triplets to longer vertical sequences, unfolding both upwards and downwards, marking consecutive classes that always lie at perceptual distance $1$ from their neighbors. The thresholds $\varepsilon_{II}(\bm{x}, \bm{b}, \hat{\bm{e}}^i)$ would be represented by the vertical separation of consecutive dots. Linearly growing thresholds, as those of Fig.~\ref{f5}A, imply that classes become increasingly separated as we depart from the diagonal. Yet, in {\sl Experiment II}, the range of colors was restricted by the gamut of the computer monitor, so the achievable chromatic difference between stimulus and surround was limited. Hence, the linear relation could only be confirmed for the limited range around the diagonal, where $\varepsilon_{II}$ is well approximated by a linear function of its arguments. When defining the perceptual coordinates, we guaranteed that classes were equi-distant right on the diagonal. Yet, beyond the diagonal, in principle distances could vary. {\sl Experiment II} showed that the separation  $\varepsilon_{II}(\bm{x}, \bm{b}, \hat{\bm{e}}^i)$ depended only on the distance $|x_i' - b_i'|$.  Therefore, if the distance $|x_i' - b_i'|$ is changed in a fixed amount, the separation is always the same, irrespective of the individual values of $x_i'$ and $b_i'$. At least in some region around the diagonal, the lines representing the classes are rigid translations one from each other. In this region, the results of {\sl Experiment II} support the isotropy and homogeneity hypothesis. 

The linear dependency of $\varepsilon_{II}(d)$ with $d$ found in {\sl Experiment II} restricts the set of feasible functions $t(d)$. For example, in the two upper panels of Fig.~\ref{f3}{\sl E}, the separation between consecutive lines is constant, so the results of {\sl Experiment II} discard these two options. The two lower panels correspond to cases in which $\varepsilon_{II}(d) - 1 \propto  d$, for small $d$. Therefore, thus far, they both constitute possible candidate descriptions of the effect of the surround on the classes of equivalence. We now compare these options.

Let us first assume that the initial linear trend apparent in the data shown in Fig.~\ref{f5} continues also for larger distances. This hypothesis implies that $\varepsilon_{II}$ is proportional to $ 1 + \lambda d(\bm{x}, \bm{b})$. It then follows that $t'(d) / t'(0) = (1 - \lambda d)^{-1}$, which in turn yields $t(d) = t'(0)  \ \ln(1 + \lambda d)/  \lambda$. The resulting displacement $t(d) - d$ is illustrated in panel {\sl d1} of Fig.~\ref{f0002}. The effect of the surround is initially repulsive, becomes neutral at an intermediate distance in which $t(d) = d$, and reverts to attractive for even larger distances (see the inversion of the arrows representing the vector field in panel {\sl c2} of Fig.~\ref{f0001}). Actually, $t(d)$ can even become negative. This behavior challenges our intuition in several ways, namely:
\begin{enumerate}
    \item[-] Thresholds grow unbounded, implying that sufficiently distant surrounds preclude the discrimination of stimuli altogether, no matter how different.
    \item[-] The displacement induced by the surround grows indefinitely for large distances. Therefore, the shifted color may differ from the presented one in an arbitrary amount, by simply displacing the surround far enough.
    \item[-] The effect inverts its polarity (from repulsive to attractive) as the distance grows. The distance where the inversion takes place is singled out. 
    \item[-] Two different surrounds (one on each side of the neutral point) acting on the same stimulus may induce the same apparent color, even though intermediate surrounds produce different apparent colors.
    \item[-] If the distance between the stimulus and the surround is sufficiently large, $t(d)$ vanishes. At that point, the stimulus becomes equal to the surround, producing a spatially uniform percept. At even larger distances, the perceived stimulus is on the negative side of the geodesic. In other words, a green stimulus surrounded by red can give rise to a red percept that is even more saturated than the surround.
\end{enumerate}
In order to avoid these bizarre effects, thresholds should deviate from the linear behavior at large distances, decelerating. The simplest deviation from the linear hypothesis would be for thresholds to saturate after the initial linear growth. Such saturation can be modelled as $\varepsilon_{II}(d) \propto [1 + a \exp(- d / \lambda)]^{-1}$, as in panel {\sl e1} of Fig.~\ref{f0002}. The limited range in which {\sl Experiment II} was performed (Fig.~\ref{f5}) does not show strong evidence of saturation. Yet, one can still test whether the thresholds of Fig.~\ref{f5} can also be compatible with a sublinear trend. To this end, we compared the hypotheses that $t'(d) \propto (1 + d / \lambda)^{-1}$ (compatible with linear thresholds) and $t'(d) \propto 1 + a\mathrm{e}^{-d/\lambda}$ (compatible with exponentially saturating thresholds). Slightly smaller  $\chi^2$ values were obtained for the exponential model. Even though the improvement in the fit of {\sl Experiment II} was only marginal, 
in the next section we describe {\sl Experiment III} with the exponentially saturating model, thereby avoiding the unrealistic effects described above.


\subsection{Experiment III: Asymmetric matching task}
\label{s:exp3}

In the asymmetric matching task (Sect.~\ref{sect:asymmetric}), for each stimulus-surround pair $\bm{x}^\alpha \sslash \bm{b}^\alpha$ and surround $\bm{b}^\beta$ the task of the observer was to find the stimulus $\bm{x}^\beta$ that fulfills $\bm{x}^\alpha \sslash \bm{b}^\alpha \sim \bm{x}^\beta \sslash \bm{b}^\beta$, in other words, to report $\bm{x}^\beta = \bm{\Phi}_{\bm{b}^\alpha \to \bm{b}^\beta}(\bm{x}^\alpha)$.

\begin{figure}[ht]
\centering
\includegraphics[scale=1]{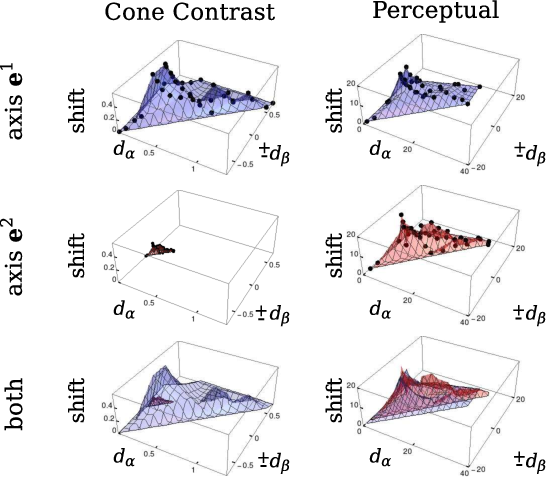}
\caption{{\bf Perceptual shifts induced by surrounds}. Shifts as a function of the distances $d^\alpha = |x_i^\alpha - b_i^\alpha|$ and $\pm d^\beta = \pm |x_i^\beta - b_i^\beta|$, for observer S3, along the axis $\hat{\bm{e}}^1$ (top), $\hat{\bm{e}}^2$ (middle) and both together (bottom), in cone contrast coordinates (left) and perceptual coordinates (right).  The factor $\pm 1$ multiplying $d^\beta$ is defined by the product of $\mathrm{Sign}$ functions in Eq.~\ref{e:exp3manifold}. The measured data points appear in the top and middle panels, and the surface interpolates the measured values. In the lower panels, the two sheets are shown to coalesce in the perceptual coordinates, but not in the cone contrast. \label{f:resexp3a}}
\label{f6} 
\end{figure}

Equation~\ref{e:dosfis} implies that this condition is equivalent to \begin{equation} \label{e:coinciden}
\bm{\Phi}_{\bm{b}^\beta}(\bm{x}^\beta) = \bm{\Phi}_{\bm{b}^\alpha}(\bm{x}^\alpha).
\end{equation}
In the following, all calculations are performed in the perceptual coordinates, but the prime symbols will be omitted to avoid cumbersome notation.  If the stimulus and the surround are both on the same cardinal axis $\hat{\bm{e}}^i$, Eq.~\ref{eq:dispnat} yields
\begin{equation} \label{e:phicordsnormales}
\left.\bm{\Phi}_{\bm{b}}(\bm{x})\right|_i = \bm{\gamma}\left( t\left(d(\bm{x}, \bm{b})\right)\right)_i = b_i + t[d(\bm{x}, \bm{b})] \ \mathrm{Sgn}[x_i - b_i].
\end{equation}
If this condition is inserted in Eq.~\ref{e:coinciden}, 
\[
b^\beta_i + t[d(\bm{x}^\beta, \bm{b}^\beta)]\ \mathrm{Sgn}[x^\beta_i - b^\beta_i] = b^\alpha_i + t[d^\alpha(\bm{x}, \bm{b}^\alpha)] \ \mathrm{Sgn}[x^\alpha_i - b^\alpha_i].
\]
In the perceptual coordinates, $d(\bm{x}, \bm{b}) = \left|x_i - b_i\right|$. Using this equality, and a few algebraic manipulations,
\begin{equation}
 \left| \left[t(d^\beta) - d^\beta\right]  - \left[ t(d^\alpha) - d^\alpha \right] \ \mathrm{Sgn}[x^\alpha_i - b^\alpha_i] \ \mathrm{Sgn}[x^\beta_i - b^\beta_i] \right| = \left| x^\alpha_i - x^\beta_i \right| \label{e:exp3manifold}
\end{equation}
Therefore, the perceptual shift $\left| x_i^\beta - x_i^\alpha \right|$ induced by the two surrounds only depends on the distances $d^\alpha = d(\bm{x}^\alpha, \bm{b}^\alpha)$ and $d^\beta = d(\bm{x}^\beta, \bm{b}^\beta)$ between  each stimulus and its surround: As long as $d^\alpha$ and $d^\beta$ remain constant, the shifts depend on none of the individual values $x_i^\alpha, x_i^\beta, b_i^\alpha$ or $b_i^\beta$, nor on the direction $\hat{\bm{e}}^i$. As a consequence, if shifts are plotted as a function of $d^\alpha$ and $d^\beta$, the set of data points should define a $2$-dimensional manifold, no matter how many stimuli, surrounds and cardinal axes be included. Moreover, the $2$-dimensional structure should only be evident in the perceptual coordinates, since in any other coordinate system, $d \ne \left|x_i - b_i \right|$, implying that Eq.~\ref{e:exp3manifold} does not hold. In Fig.~\ref{f:resexp3a}, the obtained graphs are displayed.
Along each coordinate axes, the shifts define a $2$-dimensional manifold, both in the cone contrast and the perceptual coordinates. If both axes are mixed, however, in the perceptual coordinates the collection of data points still lie on a $2$-dimensional manifold, since the two sheets corresponding to the different axes coalesce. This is not the case in the cone contrast coordinates, since the sheet corresponding to $\hat{\bm{e}_2}$ is significantly closer to the origin than that of $\hat{\bm{e}_1}$. To quantify this difference, we estimated the dimension $D$ of the manifold containing the data \cite{Granata2016}, obtaining $D = 2.11$ in the perceptual coordinates, and $D = 3.19$ in the cone contrast coordinates. 

\begin{figure}[ht]
\centering
\includegraphics[width=\textwidth, keepaspectratio]{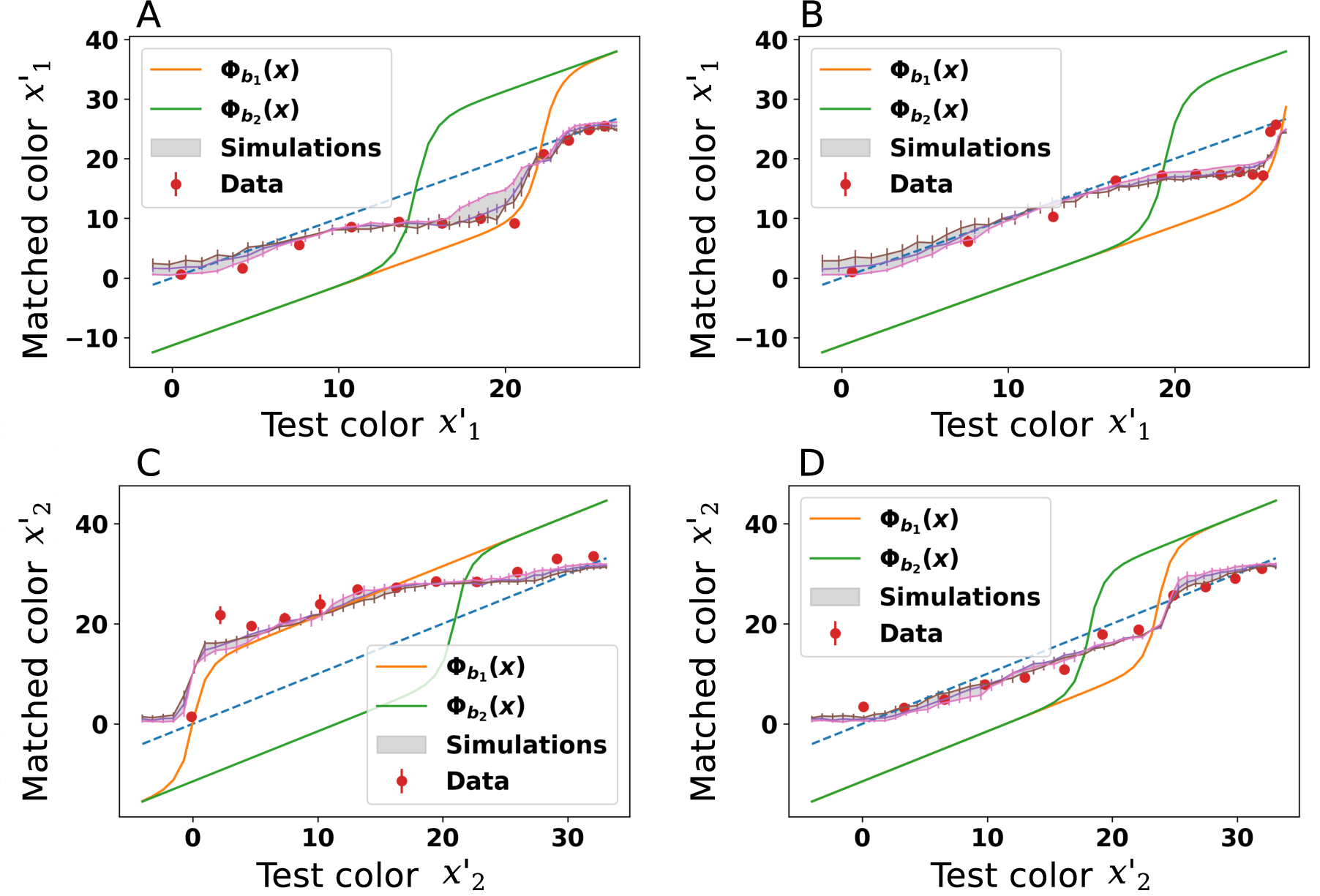}
\caption{{\bf Comparison between the measured and simulated data in {\sl Experiment III}}. Matched chromaticity $\bm{x}^\beta$ as a function of the target chromaticity $\bm{x}^\alpha$ (both axes in the perceptual coordinates) for observer S3. {\sl A} and {\sl B}: Asymmetric matching along the axis $\hat{\bm{e}}^1$. {\sl C} and {\sl D}: Asymmetric matching along the axis $\hat{\bm{e}}^2$. In the cone contrast coordinates, the two surrounds were $\bm{b}^\alpha = (-0.35, 0)$ and $\bm{b}^\beta = (0.25, 0)$ ({\sl A}), $\bm{b}^\alpha = (0, 0)$ and $\bm{b}^\beta = (0.9, 0)$ ({\sl B}); $\bm{b}^\alpha = (0, 0)$ and $\bm{b}^\beta = (0, -0.2)$ ({\sl C}), $\bm{b}^\alpha =(0, -0.03)$ and $\bm{b}^\beta = (0, 0.03)$ ({\sl D}). Red circles: experimental data. Violet line: Simulated results.  
Shaded areas: standard deviation of the simulated results. Blue dotted line: identity function, expected in the case in which the surround exerts no influence. Green and orange lines: mappings $\bm{\Phi}_{\bm{b}^\alpha}(\bm{x}^\alpha)$ and $\bm{\Phi}_{\bm{b}^\beta(\bm{x}^\alpha)}$ obtained from the fitted values of $\kappa$ and $\lambda$ of observer S3, indicating the uniform representatives of $\bm{x}^\alpha \sslash \bm{b}^\alpha$ and $\bm{x}^\alpha \sslash \bm{b}^\beta$, respectively. The perceptual shift induced by the surround becomes relevant in the interval of $\bm{x}^\alpha$ values for which the two shifts (green and orange curves) are unequal, thereby producing a net unbalance. \label{f:resexp3b}}
\end{figure}

In order to test whether the exponential model provided a good description of the results of {\sl Experiment III}, for each human observer we simulated a computational agent performing the same forced choice task. The agent decided in each trial which of the two candidate stimuli was most similar to the target, and did so according to their own idiosyncratic metric, as determined in {\sl Experiment I}. This metric was used to represent the experiment in the perceptual coordinates. In these coordinates, the effect of the surround was modeled as $\left.\bm{\Phi}_{\bm b}(\bm{x})\right|_i =  x_i + \kappa \ \mathrm{Sign}(x_i - b_i) \ [1 - \exp(- \vert x_i - b_i \vert/\lambda)]$. This functional form gives rise to an initial linear growth of $\epsilon_{II}(d)$, and an exponential saturation for long distances. For each target color $\bm{x}^{\alpha}$ presented on a surround $\bm{b}^{\alpha}$ and two candidate chromaticities $\bm{x}^p$ and $\bm{x}^q$ on the surround $\bm{b}^\beta$, the agent had to decide whether $d[\bm{\Phi}_{\bm{b}^{\alpha}}(\bm{x}^{\alpha}), \bm{\Phi}_{\bm{b}^\beta} (\bm{x}^p)]$ was larger or smaller than $d[\bm{\Phi}_{\bm{b}^{\alpha}}(\bm{x}^{\alpha}), \bm{\Phi}_{\bm{b}^\beta} (\bm{x}^q)]$. Guided by the choices of the agent, the iterative procedure of the experiment produced the final $\bm{x}^\beta = \bm{\Phi}_{\bm{b}^\alpha \to \bm{b}^\beta}(\bm{x}^\alpha)$. Since significant amounts of trial-to-trial variability were observed in the responses (Fig.~\ref{f3}), additive Gaussian noise, with zero mean and a variance fitted for each subject, was included in the evaluation of the distances $d[\bm{\Phi}_{\bm{b}^{\alpha}}(\bm{x}^{\alpha}), \bm{\Phi}_{\bm{b}^{\beta}}(\bm{x} ^p)] $ and $d[\bm{\Phi}_{\bm{b}^{\alpha}} (\bm{x}^{\alpha}),  \bm{\Phi}_{\bm{b}^\beta}(\bm{x}^q)]$ computed by the simulated observers. The simulated responses were therefore also stochastic. The functional form proposed for $t(d)$ contains two free parameters, $\kappa$ and $\lambda$. The fitting procedure was implemented with the python package \emph{noisyopt} [\cite{Spall1998}, \cite{Mayer2016}], which handles noisy functions. A single exponential function and a single noise variance was fitted for each observer, for the three different pairs of surrounds on each axis, and for both axes. Figure \ref{f:resexp3b} displays the resulting $\bm{x} ^\beta$ values as a function of the target $\bm{x}^\alpha$ for subject S3, on four different pairs of surrounds, two for each axis.  
Simulated matches may sometimes appear to be discontinuous (for example first data point in Fig.~\ref{f:resexp3b}C). This behavior derives from the staircase procedure employed to approach the matched stimulus, since for some  test chromaticities, neither the human nor the simulated subjects can select chromaticities that (in their subjective experience) are shifted by the induction of the surround outside the two offered options. The shift induced by the surround becomes significant in the interval of target $\bm{x}^\alpha$ values in which the push/pull produced by one of the surrounds is not compensated by the other, that is, where the green and orange lines differ. The simulations reproduced qualitatively the measured data. 


\section{Discussion}
\label{sec:conclu}

This paper constructs a notion of distance from a Riemannian geometry in a perceptual space, such that the symmetries governing the mapping of sensory stimuli to percepts
are most simply revealed. It does so in color space, as an example of a perceptual space in which the discriminability of neighboring stimuli does not depend linearly on notions of distance defined in terms of simple quantities derived from the physical stimulus. Our work embraces the conceptual framework first introduced by \cite{Resnikoff1974}, and recently reviewed by \cite{Provenzi2020}, in which color is understood as a property of classes of equivalence in the space of  stimulus-surround pairs. This framework was based on the observation that colored surrounds modify the appearance of chromatic stimuli. Our starting point was the assumption that, far away from the borders of color space, the perceptual effect of a given surround on a given stimulus is governed by a universal law. Here, ``universal'' means that a notion of distance $d$ between classes exists, such that  
\begin{equation} \label{eq:postulado}
\bm{\Phi}_{\bm{b}}(\bm{x}) = \bm{\gamma}_{\bm{b} \to \bm{x}}\{t[d(\bm{x}, \bm{b})]\},
\end{equation}
where $\bm{\gamma}_{\bm{b} \to \bm{x}}$ is the geodesic connecting $\bm{b}$ and $\bm{x}$ obtained from the postulated distance, and $t$ is some function that we still need to specify. Equation~\ref{eq:postulado} is a strong assumption. If no symmetries are assumed, $\bm{\Phi}_{\bm{b}}(\bm{x})$ can be any transformation $\mathbb{R}^3 \times \mathbb{R}^3 \to \mathbb{R}^3$. Once Eq.~\ref{eq:postulado} is imposed, the characterization of $\bm{\Phi}_{\bm{b}}(\bm{x})$ reduces to determining the function $t: \mathbb{R}^+ \to \mathbb{R}^+$, which is a much simpler object. If, in addition, the postulated distance derives from a metric tensor with zero curvature, then a system of coordinates exists, here called the {\sl perceptual coordinates}, in which the perceptual distance is Euclidean. In this coordinate system, all the classes of equivalence have the same shape, and only differ from one another by a rigid translation. The freedom in the shape of $t(d)$ implies that there is freedom in the shape of a single class. Yet, once the manifold corresponding to a single class is known, all others are known too.

In this paper, we tested the hypothesis that the notion of distance required to model chromatic induction through Eq.~\ref{eq:postulado} also governed the 
similarity of chromatic stimuli in terms of discrimination thresholds. If a single notion of distance is involved in a variety of experiments, one may suspect that the space of colors indeed possesses a natural geometry, accessible by many - if not all - the computations implicated in the transformation from input stimuli into behavioral responses. It therefore makes sense to study the geometry of color space, because such geometry is not idiosyncratic to specific tasks: It remains invariant throughout a variety of paradigms. The invariance suggests that the geometry is more a property of the phenomenal experience of color per se, and not of the requested behavior.

Previous work \cite{Krauskopf1992, Fonseca2016, Fonseca2018} had demonstrated that, when color was represented in the cone contrast coordinates, the principal axes of the discriminationcontain ellipsoids were aligned with the coordinate axes. In these coordinates, the metric tensor is decomposable as a direct sum and therefore, has zero curvature. 
It should be mentioned, however, that other previous studies exist in which the curvature was assumed to be negative, or to change sign throughout color space. For example \cite{Resnikoff1974}, who was later followed by \cite{Lenz2007} and \cite{Farup2014}, had theoretical reasons to consider the hypothesis of an hyperbolic geometry. His theory was based on the premise that the space of colors had to be homogeneous with respect to the general linear group of transformations, a conjecture that inspired additional studies \cite{Berthier2019, Berthier2020, Berthier2021, Berthier2021b}. Resnikoff's mathematical analysis demonstrated that, as a consequence, the classes of equivalence are forced to be linear, that is, subspaces of the $6$-dimensional space of stimuli $\times$ surrounds, as sketched in Fig.~\ref{f3}B. Those classes permit only two geometries: hyperbolic (negative curvature) and flat (null curvature). Yet, Resnikoff's conjecture still needed experimental verification. Linear classes predict constant thresholds for {\sl Experiment II}, which are clearly refuted by Figs.~\ref{f4} and \ref{f5}. They also predict a linear matching function for {\sl Experiment III}, which is refuted by Fig.~\ref{f:resexp3b}. Our experiments, hence, rebut the homogeneity hypothesis proposed by Resnikoff, thereby responding to the query raised by \cite{Provenzi2020} and \cite{Provenzi2021}. Yet, our proposal can be understood as a generalization of Resnikoff's ideas. His homogeneity hypothesis claimed the space of colors to have a specific symmetry, that allowed him to drastically reduce the range of possible structures that the space could be endowed with. We have disproved the linear structure, thereby rejecting his specific choice of symmetry. Yet, we still claim the induction to take the simplest form that is compatible with the metric determined by discrimination experiments, which is to be radial, isotropic and homogeneous.

Two other studies \cite{Silberstein1943, Kohei2011} considered other types of curvatures. By interpolating the discrete set of points measured by  \cite{MacAdam1942} with  continuous quadratic forms, they derived a metric whose curvatures changed sign throughout color space. The curvature tensor is obtained from the second derivatives of the metric, which in turn, depends on the fitted ellipses. We have verified that a small amount of variability in the measured ellipses easily modifies the sign of the curvature (data not shown), although their absolute values typically remain small. In this context, we have here used as a starting point the ellipses measured by \cite{Krauskopf1992}, which are compatible with a vanishing curvature tensor, as also argued theoretically by \cite{Fonseca2018}. This premise can be taken as an approximation that holds in the vicinity of the reference gray used in our experiments. We remain open to the existence of a small, non-vanishing curvature that may be confirmed by future, more precise experiments, particularly if the exploration of the space of colors is extended to include more saturated stimuli.

The curvature is an invariant property that does not depend on the coordinates. Hence, the hypothesis of vanishing curvature implies that the perceptual coordinates exist.
Importantly, in this paper we concluded that the transformation yielding the perceptual coordinates was significantly different for different observers, implying that no unique coordinate system exists that is perceptually uniform for all trichromats. This finding is in line with the subject-to-subject variability obtained in theoretical \cite{Fonseca2016} and experimental \cite{Wyszecki1971, Webster2002} studies of discrimination tasks, the population variability in color matching experiments \cite{Stiles1959, Wyszecki1971, Alfvin1997, Fairchild2013, Fairchild2016, Asano2016a, Asano2016b, Fonseca2021}, and experimental studies on chromatic memory \cite{Fonseca2019}. It is also consistent with the recurrently failed attempts to define a unique coordinate system perceived as perceptually uniform by all observers.

The metric was defined from discrimination thresholds measured around the uniform condition ({\sl Experiment I}). In order to verify whether the homogeneity and isotropy hypotheses entailed in Eq.~\ref{eq:postulado} hold, we performed {\sl Experiments II} and {\sl III}, and described them in terms of the metric obtained from {\sl Experiment I}, with no fitted parameters. The last two experiments, however, required in addition the function $t(d)$.

{\sl Experiment II} was restricted to regions of color space in which $\bm{x}$ remained fairly close to $\bm{b}$, due to the limited range of colors that can be produced by a computer monitor. Since thresholds grow as the chromatic distance between stimulus and surround increases, the range of discrimination experiments that can be performed with contrasting surrounds is limited. Therefore, only the first order Taylor expansion of $t(d)$ could be obtained from {\sl Experiment II}. That first order confirmed that thresholds did not remain constant when the distance between stimulus and surround was varied, thereby contradicting Resnikoff's conjecture of linear classes. In addition, the validity of Eq.~\ref{eq:postulado} was corroborated (Fig.~\ref{f5}). 

The limited range explored by {\sl Experiment II} was  overcome by {\sl Experiment III}, in which a perceptual match, instead of a discrimination, was required from the observer. The larger range of explored distances implies that now the full $t(d)$ is required to describe the experiment. Once the hypothesis of constant thresholds is discarded, the one that follows in simplicity assumes that thresholds grow linearly with $d$. Yet, this assumption implies that $t(d)$ grows in a logarithmic manner, which means that the shift $t(d) - d$ produced by the surround changes sign, a behavior that is counterintuitive. The simplest next alternative is that after an initial linear trend, thresholds decelerate, and do so sufficiently fast so as to force the perceptual shift $t(d) - d$ to converge towards a constant value for large distances. One simple way to model this behavior is with thresholds that approach exponentially an upper bound.   This model again confirmed the validity of Eq.~\ref{eq:postulado} (Fig.~\ref{f6} ) and was able to reproduce the temporal sequence of choices of subjects, as illustrated in Fig.~\ref{f:resexp3b}.

The universality entailed in Eq.~\ref{eq:postulado} suggests that the same mechanism by which surround $\bm{b}_1$ modifies the color of stimulus $\bm{x}_1$ is active when surround $\bm{b}_2$ modifies the color of stimulus $\bm{x}_2$. This mechanism is likely to be implemented by lateral or convergent feedforward connections underlying modulatory interactions in visual neurons \cite{Zeki1983,Schein1990,Wachtler2003} and may be the same mechanism underlying perceptual shifts in different modalities \cite{Klauke2015}. If a single physiological mechanism is responsible for the induction observed in different regions of color space, then the perceptual coordinates are probably the substrate upon which the synaptic processes instantiating induction operate. This hypothesis would imply that the perceptual coordinates represent signals that actually exist in the brain, and not just a mathematical construct. 

The conclusions supported by our experiments can only be claimed to hold far away from the borders of color space, since this is the region that could be tested with our computer monitor. Color space is confined into a cone included inside the positive portion of the $3$-dimensional $SML$ space, the borders of which are the maximally saturated colors. These colors cannot be generated with broadband stimuli as produced by computer displays. The existence of a border in color space blatantly contradicts the homogeneity hypothesis. We therefore take special care to limit the validity of our results, since color space cannot be homogeneous near its borders. As a consequence, the exponential model for the repulsive effect produced by surrounds cannot hold near maximally saturated stimuli, since it would push the color outside the boundaries of the space. Maybe, close to the borders, chromatic induction diminishes. Physiologically, this would mean that when color-representing neurons are firing within a certain specific range (probably their maximal rates) the synaptic mechanisms mediating the chromatic induction produced by surrounds becomes negligible. Another possibility is that chromatic induction remains constant, but that the metric becomes singular near the borders. If thresholds tend to zero sufficiently fast as we approach saturated colors, in perceptual coordinates the border of color space would be pushed away to infinity. A third alternative would be that chromatic induction still holds at the border, so that  colors are indeed pushed outside the space generated by uniform representatives. This would imply that the assumption that all classes of equivalence contain a uniform representative breaks down at the borders of color space, and the most saturated colors from the perceptual point of view always correspond to  non-matching stimulus and surround. This would be compatible with the claimed existence of the so called chimerical/hyperbolic colors \cite{Churchland2005}. New experiments with saturated colors are required to differentiate these alternatives.

We conclude that the space of colors can be endowed with a notion of distance and a system of coordinates that transparently reveal the symmetry of perceptual effects. The notion of distance stems from discrimination experiments in which just-noticeable differences are used to define the metric tensor. We hope these results motivate similar attempts in other perceptual spaces and other sensory modalities, so that the generality of these results can be assessed.


\begin{supplement}
	\stitle{Experimental data availability}
	\sdescription{Experimental data can be found at 
		\url{https://doi.org/10.12751/g-node.cwbvw6}.}
\end{supplement}

\begin{supplement}
\stitle{Appendix}
\sdescription{Tables with  fitted parameters and tests of goodness of the fitted 
curves among the article.}
\end{supplement}

\begin{table}[h]
\caption{\label{t1} Parameters of the linear and quadratic fits of $x'_i(x_i)$. The reported $p$-values represent the probability that data as extreme as the ones obtained in the experiment be generated with the fitted model. }
\centering
\small
\begin{tabular}{cccc|ccccc}
\hline
\multicolumn{4}{c}{$J_{11}$} & \multicolumn{5}{c}{$J_{22}$} \\ \hline
Sub. &             $\alpha_0$ &              $\alpha_1$ &       $p$-value &  Sub. & $\alpha_0 (. 10^2)$ & $\alpha_1 (.10^2)$ & $\alpha_2 (.10^3)$ &  $p$-value \\
\hline
\centering
S1 &  $51 \pm 3$ &  $-47 \pm 8$ &  $0.9962$ &  S1 &  $2.14 \pm 0.13$ &  $-2.0\pm 0.4$ &  $-1.7 \pm 0.8$ &  $0.8826$ \\ 
S2 &  $39 \pm 3$ &  $-33 \pm 6$ &  $0.9995$ &  S2 &  $3.4 \pm 0.2$ & $-0.55 \pm 0.35$ &  $-1.7 \pm 0.8$ &  $0.9991$ \\
S3 &  $36 \pm 2$ &  $-42 \pm 7$ &  $0.9933$ &  S3 &  $1.85 \pm 0.08$ &  $-2.9\pm 0.3$ &  $-0.94 \pm 0.44$ &  $0.4808$ \\
S4 &  $38 \pm 3$ &  $-24 \pm 7$ &  $0.2271$ &  S6 &  $1.72 \pm 0.08$ &  $-1.1 \pm 0.2$ &   $-1.7 \pm 0.5$ &  $0.5026$  \\
S5 &  $31 \pm 2$ &  $-17 \pm 6$ &  $0.6555$ &  S7 &  $1.89 \pm 0.14$ &  $-2.2 \pm 0.4$ &  $-2.0 \pm 0.8$ &  $0.9901$ \\
\hline
\end{tabular}
\end{table}
%

\begin{table}[h]
\footnotesize
\caption{Fitted coefficients for Models 1 and 2 (Eqs.~\ref{e:exp2mod1} and \ref{e:exp2mod2}) for all measured subjects along the axis $\hat{\bm{e}}^1$.}
\label{t:e2ejeS}
\centering
\scalebox{1}{
\begin{tabular}{|c | c c c | c c c |}
\toprule
  {}& {}&  Model 1     &  {}&{}& Model 2&{}\\ \midrule
{} & $\gamma_{0} $ & $\gamma_1$ & $\gamma_2$ & $\gamma_0$ &  $\gamma_1$ & $\gamma_2$ \\
\midrule
S1 &   $0.014 \pm 0.004$ &   $0.058 \pm 0.01$ &  $0.21 \pm 0.04$ &    $0.001 \pm 0.005$ &   $0.13 \pm 0.02$ &   $0.053 \pm 0.01$  \\
S2 &  $0.0034 \pm 0.002$ &  $0.032 \pm 0.009$ &  $0.24 \pm 0.02$ &     $-0.015 \pm 0.003$ &   $0.18 \pm 0.01$ &  $0.046 \pm 0.009$ \\
S3 &  $0.0071 \pm 0.004$ &   $0.047 \pm 0.01$ &  $0.11 \pm 0.03$ &     $-0.0043 \pm 0.005$ &  $0.087 \pm 0.02$ &  $0.043 \pm 0.009$ \\
S4 &  $0.0075 \pm 0.003$ &   $0.019 \pm 0.01$ &  $0.23 \pm 0.03$ &     $-0.0036 \pm 0.004$ &   $0.13 \pm 0.02$ &   $0.012 \pm 0.01$ \\
S5 &  $0.0075 \pm 0.003$ &   $0.055 \pm 0.01$ &  $0.33 \pm 0.03$ &     $-0.012 \pm 0.004$ &    $0.2 \pm 0.02$ &   $0.054 \pm 0.01$ \\
\bottomrule
\end{tabular}}
\end{table}

\begin{table}[h]
\footnotesize
\caption{Fitted coefficients for Models 1 and 2 (Eqs.~\ref{e:exp2mod1} and \ref{e:exp2mod2}) for all measured subjects along the axis $\hat{\bm{e}}^2$.}
\label{t:e2ejeLM}
\centering
\scalebox{1}{
\begin{tabular}{|c | c c c | c c c|}
\toprule
  {}& {}&  Model 1     &  {}&{}& Model 2&{} \\ \midrule
{} & $\gamma_0$ & $\gamma_1$ & $\gamma_2$ & $\gamma_0$ & $\gamma_1$ & $\gamma_2$ \\
\midrule
S1 &  $0.0031 \pm 0.0008$ &   $0.018 \pm 0.02$ &   $0.73 \pm 0.1$ &  $7.3e-06 \pm 0.001$ &  $0.13 \pm 0.02$ &   $0.018 \pm 0.01$  \\
S2 &  $0.0028 \pm 0.0003$ &   $0.02 \pm 0.007$ &   $0.8 \pm 0.07$ &     $-0.00015 \pm 0.0004$ &  $0.13 \pm 0.01$ &   $0.02 \pm 0.007$  \\
S3 &  $0.0022 \pm 0.0007$ &  $0.024 \pm 0.009$ &  $0.69 \pm 0.09$ &       $-0.0011 \pm 0.001$ &  $0.12 \pm 0.01$ &   $0.02 \pm 0.009$  \\
S6 &   $0.0096 \pm 0.002$ &   $0.051 \pm 0.02$ &      $1 \pm 0.2$ &       $0.0037 \pm 0.002$ &  $0.19 \pm 0.03$ &   $0.039 \pm 0.02$  \\
S7 &  $0.0024 \pm 0.0004$ &  $0.048 \pm 0.008$ &     $1 \pm 0.08$ &    $-0.0013 \pm 0.0006$ &  $0.16 \pm 0.01$ &  $0.047 \pm 0.008$ \\
\bottomrule
\end{tabular}}
\end{table}





\vspace{1cm}

\providecommand{\bysame}{\leavevmode\hbox to3em{\hrulefill}\thinspace}
\providecommand{\MR}{\relax\ifhmode\unskip\space\fi MR }
\providecommand{\MRhref}[2]{%
  \href{http://www.ams.org/mathscinet-getitem?mr=#1}{#2}
}
\providecommand{\href}[2]{#2}


\ACKNO{This work was supported by the Agencia Nacional de Promoción de la 
	Investigación, el Desarrollo Tecnológico y la Innovación, the Consejo Nacional de 
	Investigaciones Científicas y Técnicas, the Comisión Nacional de Energía Atómica 
	and the Universidad Nacional de Cuyo of Argentina, and by the Bernstein Center 
	for 
	Computational Neuroscience Munich, Germany. The authors wish to thank the 
	subjects that participated in the experiment, and to Sarah Theimer and Hongbin 
	Wu 
	for summoning volunteers and discussing previous versions of the behavioral 
	paradigm.
}

\nocite{*}
\end{document}